\documentclass[aps,prd,reprint,superscriptaddress,showkeys,amsmath,amssymb,amsfonts]{revtex4-1}
\usepackage[T1]{fontenc}
\usepackage[utf8]{inputenc}
\usepackage[spanish,USenglish]{babel}
\usepackage[colorlinks,citecolor=blue,urlcolor=blue,linkcolor=blue]{hyperref}
\usepackage{physics}
\usepackage{enumerate}
\usepackage{graphicx}
\usepackage{dcolumn}
\usepackage{multirow}

\newcommand{\schr}{Schr\"{o}dinger}
\newcommand{\BO}{B\nobreakdash-O}

\begin{document}

\title{Diabatic description of charmonium-like mesons}
\author{R. Bruschini}
\email{roberto.bruschini@ific.uv.es}
\affiliation{\foreignlanguage{spanish}{Unidad Teórica, Instituto de Física Corpuscular (Universidad de Valencia -- CSIC), E-46980 Paterna (Valencia)}, Spain}
\author{P. González}
\email{pedro.gonzalez@uv.es}
\affiliation{\foreignlanguage{spanish}{Unidad Teórica, Instituto de Física Corpuscular (Universidad de Valencia -- CSIC), E-46980 Paterna (Valencia)}, Spain}
\affiliation{\foreignlanguage{spanish}{Departamento de Física Teórica, Universidad de Valencia, E-46100 Burjassot (Valencia)}, Spain}


\begin{abstract}
We apply the diabatic formalism, first introduced in molecular physics, to the description of heavy-quark mesons. In this formalism the
dynamics is completely described by a diabatic potential matrix whose elements can be derived from unquenched lattice QCD studies of string breaking.
For energies far below the lowest open flavor meson-meson threshold,
the resulting diabatic approach reduces to the well-known Born-Oppenheimer approximation
where heavy-quark meson masses correspond to energy levels in an effective quark-antiquark potential.
For energies close below or above that threshold, where the Born-Oppenheimer approximation
fails, this approach provides a set of coupled {\schr} equations incorporating meson-meson components nonperturbatively, i.e.\ beyond loop corrections.
A spectral study of heavy mesons containing $c\overline{c}$ with masses below $4.1$~GeV is carried out within this framework.
From it a unified description of conventional as well as unconventional resonances comes out.
\end{abstract}

\maketitle

\section{\label{sec1}Introduction}

The discovery of the $\chi_{c1}(3872)$ in 2003 \cite{Cho03} may be
considered as the initio of a new era in heavy-quark meson spectroscopy. This
resonance and a plethora of new states ($\psi(4260)$, $\psi(4360)$, $X(3915)$, and many others, see \cite{PDG20})
discovered since then have masses and
decay properties that do not correspond to the
conventional heavy quark ($Q$) -- heavy antiquark ($\overline{Q}$)
meson description, such as the one provided by nonrelativistic or
semirelativistic quark models that has been so successful in the past \cite{Eic80,Eic04,GI85}. A feature of any of
these unconventional states is that its mass lies close below or above the
lowest open flavor meson-meson threshold with the same quantum numbers. This
suggests a possible relevant role of open flavor meson-meson thresholds in the
explanation of the structure of the new states. As a matter of fact, the
nonrelativistic Cornell quark model \cite{Eic80,Eic04} incorporates some of
these effects through meson loops where the interaction
connecting $Q\overline{Q}$ and open flavor meson-meson is derived from the $Q\overline{Q}$
binding potential. Similar kind of loop contributions, with quark pair creation models like
the $^{3}\!P_{0}$ one providing the valence-continuum coupling, have been extensively
studied in the literature (see for instance \cite{Bar08} and \cite{Fer19}). However,
these perturbative loop contributions seem to be insufficient for a detailed description of
the new structures. This has led to the building of phenomenological models
involving implicit or explicit meson-meson components, for example in the forms of tetraquarks, meson
molecules, and hadroquarkonium (see \cite{Ch16,Leb17,Guo17,Esp17} and references therein).

\textit{Ab initio} calculations from QCD have been also carried out. From
lattice QCD, a Born-Oppenheimer ({\BO}) approximation for heavy-quark
mesons has been developed \cite{Jug99} (for a connection with effective field
theories see \cite{Bra20} and references therein). In this approximation,
based on the large ratio of the heavy quark
mass to the QCD energy scale associated with the gluon field, the
heavy-quark meson masses correspond to energy levels of a {\schr}
equation for $Q\overline{Q}$ in an effective potential. This potential is defined by the energy
of a stationary state of light-quark and gluon fields in the presence
of static $Q$ and $\overline{Q}$ sources, which is calculated in lattice QCD.
Thus, conventional quarkonium masses are the energy levels in the ground state
potential calculated in quenched (without light quarks) lattice QCD whose form is Cornell-like \cite{Bal01}, whereas quarkonium hybrid
($Q\overline{Q}g$ bound state where $g$ stands for a gluon) masses are energy
levels in the quenched excited state potentials. Although no tetraquark potentials have
been calculated yet from lattice QCD, some information on them has been also
extracted \cite{Braa14}. The immediate question arising is whether these
hybrid and tetraquark {\BO} potentials may correctly describe or not the new
states. The answer to this question can be derived from \cite{Braa14}, where an
assignment of the masses of some of the new states to energy levels in these
potentials has been pursued. In essence, quoting this reference, although the
{\BO} approximation provides a starting point for a coherent description of the
new states based firmly on QCD, a detailed description of them
requires to go beyond quenched lattice calculations and beyond the {\BO} approximation.

An intermediate step in this direction was taken in \cite{Gon14,*Gon15,*Gon19} by
identifying the unquenched lattice energy for
static $Q$ and $\overline{Q}$ sources, when the $Q\overline{Q}$ configuration
mixes with one or two open flavor meson-meson ones \cite{Bal05,Bul19}, with a $Q\overline{Q}$ potential. This
unquenched approximation allows for some physical understanding of threshold
effects beyond hadron loops. However, the description in terms of effective
$Q\overline{Q}$ channels does not give detailed account of the configuration mixing.

In this article we take a step further to go beyond the {\BO}
approximation. For this purpose we use the diabatic approach developed
in molecular physics for tackling the configuration mixing problem (see for
instance \cite{Bae06}). This allows us to establish a general framework
for a unified description of conventional and unconventional heavy-quark meson
states. This framework is applied to the calculation of $J^{++}$ and the
low-lying $1^{--}$ meson states with $Q=c$ (charm quark) where there are
sufficient data available to test its validity.

In this manner a complete treatment of heavy-quark
meson states involving heavy quark-antiquark and meson-meson degrees of freedom,
that incorporates the results from \textit{ab initio} calculations in quenched and
unquenched lattice QCD, comes out.

The contents of the paper are organized as follows.
In Sec.~\ref{sec2} the mathematical formalism and the physical picture leading to the
{\BO} approximation for heavy-quark mesons is revisited. In Sec.~\ref{sec3} we
detail the diabatic approach and in Sec.~\ref{sec4} we adapt it to the
description of heavy-quark meson states. The application to meson states containing
$c\overline{c}$ is detailed in Sec.~\ref{sec5}. For the sake of simplicity we
consider states involving non-overlapping thresholds with small
widths. The comparison of our results to existing data serves as a stringent test
of our treatment. Finally, in Sec.~\ref{sec6} our main conclusions are summarized.

\section{\label{sec2}Born-Oppenheimer approximation in QCD}

The Born-Oppenheimer ({\BO}) approximation was developed in 1927 for the
description of molecules \cite{Bo27}, and since then it has been a fundamental
approximation in chemistry. More recently it has been employed for the
description of heavy-quark meson bound states from QCD \cite{Jug99,Braa14}.
Next, we briefly recall the main steps in its construction for the description of
a heavy-quark meson system containing a heavy quark-antiquark ($Q\overline{Q}$)
interacting with light fields (gluons and light
quarks), with Hamiltonian
\begin{equation}
H=K_{Q\overline{Q}}+H^\text{lf}_{Q\overline{Q}}
\end{equation}
where $K_{Q\overline{Q}}$ is the $Q\overline{Q}$ kinetic energy operator
\begin{equation}
K_{Q\overline{Q}}=\frac{\vb*{p}_{Q}^{2}}{2m_{Q}}+\frac
{\vb*{p}_{\overline{Q}}^{2}}{2m_{\overline{Q}}}=\frac
{\vb*{p}^{2}}{2\mu_{Q\overline{Q}}}+\frac{\vb*{P}^{2}}{2(m_{Q}+m_{\overline{Q}})}
\end{equation}
with $\mu_{Q\overline{Q}}$ being the reduced $Q\overline{Q}$ mass, 
$\vb*{p}$ ($\vb*{P}$) the $Q\overline{Q}$
relative (total) three-momentum,
and $H^\text{lf}_{Q\overline{Q}}$ the part of the Hamiltonian containing the light field energy operator and
the $Q\overline{Q}$ -- light-field interaction. Notice that $H^\text{lf}_{Q\overline{Q}}$ depends on the $Q$ and
$\overline{Q}$ positions but does not contain any derivative with respect to the $Q$ and $\overline{Q}$ coordinates.

A heavy-quark meson bound state $\ket{\psi}$ is a solution of the characteristic equation
\begin{equation}
H\ket{\psi} = E \ket{\psi}
\end{equation}
where $E$ is the energy of the state. Note that $\ket{\psi}$ contains information
on both the $Q\overline{Q}$ and light fields.

\subsection{Static limit}

The first step in building the {\BO} approximation consists in solving the dynamics of the light
fields by neglecting the $Q\overline{Q}$ motion, i.e.\ setting the kinetic energy term $K_{Q\overline{Q}}$ equal to zero.
This corresponds to the limit where $Q$ and $\overline{Q}$ are infinitely massive,
what can be justified because the $Q$ and
$\overline{Q}$ masses, $m_{Q}$ and $m_{\overline{Q}}$, are much bigger than the QCD scale $\Lambda_\text{QCD}$, which is the energy
scale associated with the light fields.

As we are interested in the internal structure of the system and this does not depend on the center of
mass motion (which coincides with the $Q\overline{Q}$ center of mass motion in
the infinite mass limit) it is convenient to use the $Q\overline{Q}$ relative
position $\vb*{r}=\vb*{r}_{Q}-\vb*{r}_{\overline{Q}}$, and work in the $Q\overline{Q}$ center of mass
frame where $\vb*{P}=0$.

In this \emph{static limit} $\vb*{r}$ is fixed, ceasing to be a dynamical variable. This is, the components of
$\vb*{r}$ can be considered as parameters, rather than operators,
in the expression of $H^\text{lf}_{Q\overline{Q}}$ that will depend operationally on the light fields
only.
We shall indicate this parametric dependence renaming $H^\text{lf}_{Q\overline{Q}}$ as $H_\text{static}^\text{lf}(\vb*
{r})$.

It is then possible to solve the dynamics of the light fields for any value of $\vb*{r}$:
\begin{equation}
(H_\text{static}^\text{lf}(\vb*{r}) -V_{i}(\vb*{r}))\ket{\zeta_{i}(\vb*{r})} =0
\label{STATIC}
\end{equation}
where $\ket{\zeta_{i}(\vb*{r})} $ are the light field eigenstates, $V_{i}(\vb*{r})$ the corresponding eigenvalues,
and $i$ stands for the set of quantum numbers labelling the eigenstates. Note that both the eigenvalues and the
eigenstates depend parametrically on $\vb*{r}$, and that for every value of $\vb*{r}$ the eigenstates
$\{\ket{\zeta_{i}(\vb*{r})}\}$ form a complete orthonormal set for the light fields:
\begin{equation}
\braket{\zeta_{j}(\vb*{r})}{\zeta_{i}(\vb*{r})} = \delta_{ji} .
\end{equation}

As for the eigenvalues $V_{i}(\vb*{r})$, they correspond to the energies of stationary states of the light fields in the
presence of static $Q$ and $\overline{Q}$ sources placed at a relative position $\vb*{r}$,
and can be calculated \textit{ab initio} in lattice QCD.

More precisely, in quenched (with gluon but not light-quark fields) lattice QCD \cite{Bal01} the ground state of the light fields
is associated with a $Q\overline{Q}$ configuration, and up to spin dependent terms that we
shall not consider the static energy of  this ground state mimics the form of
the phenomenological Cornell potential
\begin{equation}
V_\text{C}(r)=\sigma r - \frac{\chi}{r} + m_Q + m_{\overline{Q}} - \beta
\label{CPOT}
\end{equation}
with $\sigma$, $\chi$ and $\beta$ standing for the string tension,
the color coulomb strength, and a constant fixing the origin of the potential respectively.

\begin{figure}
\includegraphics{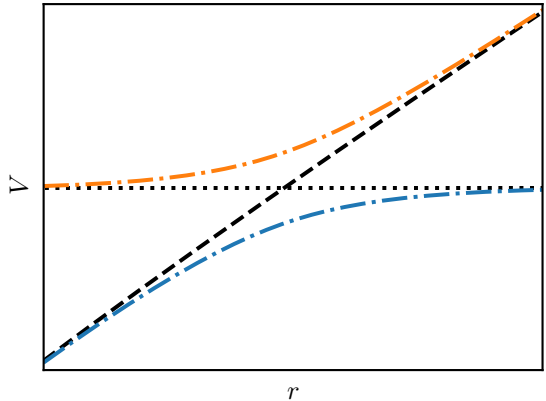}
\caption{Pictorial representation of lattice static energies. Dashed line: ground state static light field energy in quenched lattice QCD.
Dotted line: meson-meson threshold.
Dash-dotted lines: ground and excited state static light field energies in unquenched lattice QCD, showing an avoided crossing.}
\label{cross}
\end{figure}

On the other hand, unquenched (with gluon and light-quark fields) lattice QCD calculations \cite{Bal05,Bul19} have shown that due
to string breaking the association of the light field
ground state with a $Q\overline{Q}$ configuration holds only for small values of the 
relative $Q\overline{Q}$ distance $r \equiv \abs{\vb*{r}}$.
When increasing $r$ the $Q\overline{Q}$ configuration mixes significantly with meson-meson configurations.
More in detail: below (above) an
open-flavor meson-meson threshold the energy of a stationary
state of the light fields changes with $r$, from the one corresponding to the
$Q\overline{Q}$ (meson-meson) configuration to the one of meson-meson
($Q\overline{Q}$) configuration, avoiding in this manner the crossing of the static light field
energies corresponding to pure $Q\overline{Q}$ and meson-meson configurations that would take place at
the threshold mass in absence of string breaking. In Fig.~\ref{cross} we have represented graphically this situation for $Q\overline{Q}$ and one
meson-meson threshold (the representation for two meson-meson thresholds can
be seen in \cite{Bal05,Bul19}).

\subsection{Adiabatic expansion}

Having solved the static problem for the light fields, the next step in the construction of the
{\BO} approximation consists in reintroducing the $Q\overline{Q}$ motion. This is done by solving the bound state equation
\begin{equation}
\pqty{\frac{\vb*{p}^2}{2\mu_{Q\overline{Q}}} + H_\text{static}^\text{lf}(\vb*{r})-E} \ket{\psi} =0,
\label{BSECM}
\end{equation}
where $E$ denotes the mass of the bound state, making use of the so-called adiabatic expansion for $\ket{\psi}$:
\begin{equation}
\ket{\psi} =\sum_{i}\int \dd\vb*{r}^{\prime}\psi_{i}(\vb*{r}^{\prime})\ket{\vb*{r}^{\prime}}\ket{\zeta_{i}(\vb*{r}^{\prime})}
\label{BOSTATE}
\end{equation}
where $\ket{\vb*{r}^\prime}$ is a state indicating the $Q\overline{Q}$ relative position and
we have temporarily omitted spin degrees of freedom for simplicity. The qualifier ``adiabatic'' refers to the fact that each term
in the expansion depends only on a single value of $\vb*{r}^\prime$, what can be related to the physical situation where
the light fields respond almost instantaneously to the motion of the quark and antiquark. However, as will be shown in what
follows, this physical expansion is not mathematically convenient when configuration mixing takes place. Note that as the
states $\ket{\zeta_{i}(\vb*{r}^\prime)}$ depend on $\vb*{r}^\prime$, so do the coefficients $\psi_{i}$, one for each light field state.

Using \eqref{BOSTATE} and multiplying on the left by $\bra{\vb*{r}}$ the bound state equation can be rewritten as
\begin{equation}
\sum_{i}\pqty{-\frac{\hbar^{2}}{2\mu_{Q\overline{Q}}}\laplacian+V_{i}(\vb*{r})-E}\psi_{i}(
\vb*{r})\ket{\zeta_{i}(\vb*{r})} =0,
\end{equation}
then multiplying on the left by $\bra{\zeta_{j}(\vb*{r})}$ yields
\begin{widetext}
\begin{equation}
\sum_{i}\bqty{-\frac{\hbar^{2}}{2\mu_{Q\overline{Q}}}\mel{\zeta_{j}(\vb*{r})}{\laplacian\psi_{i}(\vb*{r})}{\zeta_{i}(
\vb*{r})} + \pqty{V_{j}(\vb*
{r})-E} \delta_{j i} \psi_{i}(\vb*{r})} =0.
\label{EQS}
\end{equation}
The first term on the left hand side of \eqref{EQS} can be developed as
\begin{equation}
\mel{\zeta_{j}(\vb*{r})}{\laplacian\psi_{i}(\vb*{r})}{\zeta_{i}(\vb*{r})} = \delta_{ji}\laplacian\psi_{i}(\vb*{r}) + 
2\vb*{\tau}_{ji}(\vb*{r}) \vdot \grad\psi_{i}(\vb*{r}) + \tau_{ji}^{(2)}(\vb*{r})\psi_{i}(\vb*{r})
\end{equation}
with
\begin{equation}
\vb*{\tau}_{ji}(\vb*{r}) \equiv \braket{\zeta_{j}(\vb*{r})}{\grad \zeta_{i}(\vb*{r})} \qand
\tau_{ji}^{(2)}(\vb*{r}) \equiv \braket{\zeta_{j}(\vb*{r})}{\laplacian\zeta_{i}(\vb*{r})}
\end{equation}
being the so-called \emph{Non-Adiabatic Coupling Terms} (NACTs) of the first and second order respectively.

Furthermore, using $\grad\braket{\zeta_{j}(\vb*{r})}{\zeta_{i}(\vb*{r})} =\grad\delta_{ji}=0$ we have
\begin{equation}
\vb*{\tau}_{ji}(\vb*{r})\equiv \braket{\zeta_{j}(\vb*{r})}{\grad \zeta_{i}(\vb*{r})} =
- \braket{\grad\zeta_{j}(\vb*{r})}{\zeta_{i}(\vb*{r})} \equiv-\vb*{\tau}_{ij}^*(\vb*{r}),
\end{equation}
from which it follows
\begin{equation}
\braket{\grad\zeta_{j}(\vb*{r})}{\grad\zeta_{i}(\vb*{r})}   =
\sum_{k}\braket{\grad\zeta_{j}(\vb*{r})}{\zeta_{k}(\vb*{r})} \vdot
\braket{\zeta_{k}(\vb*{r})}{\grad\zeta_{i}(\vb*{r})}
= \sum_{k}\vb*{\tau}_{kj}^*(\vb*{r})\vdot\vb*{\tau}_{ki}(\vb*{r}) = - \sum_{k}\vb*{\tau}_{jk}(\vb*{r})
\vdot\vb*{\tau}_{ki}(\vb*{r}) \equiv-(\vb*{\tau}(\vb*{r})^{2})_{ji},
\end{equation}
so that
\begin{equation}
(\grad\vb*{\tau}(\vb*{r}))_{ji} = \braket{\zeta_{j}(\vb*{r})}{\laplacian \zeta_{i}(\vb*{r})} +
\braket{\grad\zeta_{j}(\vb*{r})}{\grad\zeta_{i}(\vb*{r})}
 =\tau_{ji}^{(2)}(\vb*{r})-
(\vb*{\tau}(\vb*{r})^{2})_{ji}
\end{equation}
and finally
\begin{equation}
\mel{\zeta_{j}(\vb*{r})}{\laplacian\psi_{i}(\vb*{r})}{\zeta_{i}(\vb*{r})}  =
\delta_{ji}\laplacian\psi_{i}(\vb*{r})+2\vb*{\tau}_{ji}(\vb*{r})\vdot\grad\psi_{i}(\vb*{r})
+ ((\div\vb*{\tau}(\vb*{r}))_{ji}+(\vb*{\tau}(\vb*{r})^{2})_{ji})\psi_{i}(\vb*{r})
\equiv ((\grad+\vb*{\tau}(\vb*{r}))^2)_{ji}\psi_{i}(\vb*{r}).
\end{equation}

The bound state equation \eqref{EQS} then reads
\begin{equation}
\sum_{i}\bqty{  -\frac{\hbar^{2}}{2\mu_{Q\overline{Q}}}((\grad+\vb*{\tau}(\vb*{r}))^2)_{ji}+(V_{j}(\vb*{r})-E)\delta_{ji}}\psi_{i}(\vb*{r})=0.
\label{BSE}
\end{equation}
\end{widetext}
This is a multichannel equation where $\psi_{i}(\vb*{r})$ stands for the $i$-th component of the heavy-quark meson wave function, that is
in general a mixing of $Q\overline{Q}$ and meson-meson components. Notice though that this is not the usual {\schr} equation because of
the presence of the NACTs $\vb*{\tau}$ inside the kinetic energy operator. These terms introduce a coupling
between the wave function components and reflect the non-trivial interaction between the $Q\overline{Q}$ motion and
the light field states.

\subsection{Single channel approximation}

The last step in the construction of the {\BO} approximation consists in neglecting the NACTs inside the kinetic energy operator:
\begin{equation}
 \vb*{\tau}_{ji}(\vb*{r}) = \braket{\zeta_{j}(\vb*{r})}{\grad\zeta_{i}(\vb*{r})} \approx 0.
\label{BOCOND}
\end{equation}
This is called the \emph{single channel approximation} because the bound state
equation \eqref{BSE} then factorizes in a set of decoupled
single channel {\schr} equations
\begin{equation}
\bqty{-\frac{\hbar^{2}}{2\mu_{Q\overline{Q}}}\laplacian+(V_{j}(\vb*{r})-E)}\psi_{j}(\vb*{r})=0
\label{BOSE}
\end{equation}
where $V_{j}(\vb*{r})$, corresponding to the energy of the stationary $j$-th state
of the light fields in the presence of static $Q$ and $\overline{Q}$ sources, plays the role of an
effective potential.

Eqs.~\eqref{STATIC}, \eqref{BOSTATE}, \eqref{BOCOND} and \eqref{BOSE} define the {\BO} approximation.

Notice that the single channel approximation can be deemed reasonable only up to
$Q\overline{Q}$ distances for which the NACTs can be neglected, i.e.\ for distances where the $Q\overline{Q}$
and meson-meson configuration mixing associated with the light field eigenstates is negligible
(for a specific calculation see Sec.~\ref{angsec}).
This makes the {\BO} approximation to be justified only for bound state energies far below the lowest open flavor
meson-meson threshold. In particular, conventional heavy-quark meson masses, far below the lowest open flavor
meson-meson threshold, can be described as the energy levels in the potential
corresponding to the quenched ground state of the light fields, i.e.\ the Cornell potential.

\section{\label{sec3}Diabatic approach}

For energies close below or above an open flavor meson-meson threshold the mixing between the
$Q\overline{Q}$ and meson-meson configurations gives
rise to nonvanishing NACTs, so that the single channel approximation \eqref{BOCOND} cannot be maintained. Instead, one
has to deal with the set of coupled equations~\eqref{BSE}, which is not practicable for two reasons:
\begin{enumerate}[i)]
	\item There is no yet direct lattice QCD calculation of the NACTs $\vb*{\tau}$.
	\item When $\vb*{\tau} \ne 0$, the wave function components in the expansion \eqref{BOSTATE} do not correspond
	to pure $Q\overline{Q}$ or meson-meson but rather to a mixing of both, the amount of mixing depending on $\vb*{r}$.
\end{enumerate}

These drawbacks can be overcome through the use of the \emph{diabatic approach}, where
 one expands the bound state $\ket{\psi}$ on a basis of light field
eigenstates calculated at some fixed point $\vb*{r}_0$. As the $\Bqty{\ket{\zeta_i(\vb*{r})}}$
form a complete set for the light fields whatever the value of $\vb*{r}$, switching from a
$\Bqty{\ket{\zeta_i(\vb*{r})}}$ to $\Bqty{\ket{\zeta_i(\vb*{r}_0)}}$
is equivalent to a $\vb*{r}$-dependent change of basis in the light degrees of freedom.

The diabatic expansion of the bound state reads
\begin{equation}
\ket{\psi} =\sum_{i}\int \dd{\vb*{r}^{\prime}}\widetilde{\psi}_{i}(\vb*{r}^{\prime},\vb*{r}_{0})\ket{\vb*{r}^{\prime}} \ket{\zeta_{i}(\vb*{r}_{0})}
\label{DSTATE}
\end{equation}
where the coefficients $\widetilde{\psi}_{i}$, one coefficient for each light field state, are functions of $\vb*{r}^{\prime}$ that depend
parametrically on $\vb*{r}_{0}$.

A nice physical feature of this expansion is that the light field state $\ket{\zeta_i (\vb*{r}_0)}$ corresponding to each
component $\widetilde{\psi}_i$ does not depend on the $Q\overline{Q}$ relative position $\vb*{r}^\prime$. This means that if
one chooses the fixed point $\vb*{r}_0$ far from the avoided crossing, then the wave function components correspond to
either pure $Q\overline{Q}$ or meson-meson for any value of $\vb*{r}^\prime$. In other words, in the diabatic approach one expands the
bound states in terms of the more intuitive Fock components (pure $Q\overline{Q}$ and pure meson-meson)
instead of components which are a mixing of $Q\overline{Q}$ and meson-meson.

Substituting \eqref{DSTATE} in the bound
state equation \eqref{BSECM} and projecting on $\bra{\vb*{r}}$ yields
\begin{equation}
\sum_{i}\pqty{-\frac{\hbar^{2}}{2\mu_{Q\overline{Q}}}\laplacian+H_\text{static}^\text{lf}(\vb*{r})-E}
\widetilde{\psi}_{i}(\vb*{r},\vb*{r}_{0})
\ket{\zeta_{i}(\vb*{r}_{0})}
=0
\end{equation}
where all the derivatives are taken with respect to $\vb*{r}$.
If we now multiply on the left by $\bra{\zeta_{j}(\vb*{r}_{0})}$, as $\grad\ket{
\zeta_{i}(\vb*{r}_{0})}=0$ the equation reads
\begin{equation}
\sum_{i}\pqty{-\frac{\hbar^{2}}{2\mu_{Q\overline{Q}}}\delta_{ji} \laplacian+V_{ji}(\vb*{r},\vb*{r}_{0}) -
E\delta_{ji}}\widetilde{\psi}_{i}(\vb*{r},\vb*{r}_{0})=0
 \label{DEQ}
\end{equation}
where
\begin{equation}
V_{ji}(\vb*{r},\vb*{r}_{0})\equiv \mel{\zeta_{j}(\vb*{r}_{0})}{H_\text{static}^\text{lf}(\vb*{r})}{\zeta_{i}(\vb*{r}_{0})}
\label{DPOT}
\end{equation}
is the so-called \emph{diabatic potential matrix}.

The multichannel {\schr} equation \eqref{DEQ} together with
\eqref{DPOT} and \eqref{DSTATE} define the diabatic approach which is widely
employed in molecular physics \cite{Bae06}.

The complete equivalence between Eqs.~\eqref{BSE} and \eqref{DEQ} has been shown elsewhere \cite{Bae06}
and is reproduced, for the sake of completeness, in Appendix~\ref{apdxADT}. In short, the troublesome NACTs
in \eqref{BSE} that break the single channel approximation when configuration mixing is present
(thus invalidating the {\BO} framework) are taken into account in \eqref{DEQ}
through the diabatic potential matrix. This is utterly convenient since,
as we shall see in Sec.~\ref{mixpotsec}, the elements of this matrix are directly related to the
static light field energy levels calculated in quenched and unquenched lattice QCD.

It is also easy to show that when the single channel approximation
\eqref{BOCOND} holds the diabatic potential matrix \eqref{DPOT} becomes a diagonal
matrix containing the static light field energy levels calculated in quenched lattice QCD,
and consequently Eq.~\eqref{DEQ} reproduces the set of single channel {\schr} equations \eqref{BOSE}.

Therefore, the diabatic approach is a complete general framework appliable to conventional heavy-quark mesons
lying far below the lowest open flavor meson-meson threshold as well as to unconventional ones lying close below
or above that threshold.

\section{\label{sec4}Heavy-quark mesons in the diabatic framework}

In order to apply the diabatic framework to the description of
heavy-quark meson bound states we examine first the case of a single
meson-meson threshold. Then we proceed to the generalization to an arbitrary
number of thresholds.

\subsection{Spectroscopic equations}

Let us consider one meson-meson threshold. Let us fix a value for $\vb*{r}_0$
such that the ground state of the light fields
is associated with the $Q\overline{Q}$ configuration and the first excited
state with the meson-meson one. To make this more clear we relabel
the diabatic light field states as
\begin{equation}
\ket{\zeta_0 (\vb*{r}_0)} \rightarrow \ket*{\zeta_{Q\overline{Q}}}, \qquad
\ket{\zeta_1 (\vb*{r}_0)} \rightarrow \ket*{\zeta_{M_1\overline{M}_2}},
\end{equation}
and the diabatic wave function components as
\begin{equation}
\widetilde{\psi}_0(\vb*{r},\vb*{r}_0) \rightarrow \psi_{Q\overline{Q}}(\vb*{r}), \qquad
\widetilde{\psi}_1(\vb*{r},\vb*{r}_0) \rightarrow \psi_{M_1\overline{M}_2}(\vb*{r}) .
\end{equation}
Accordingly, we rename the diabatic potential matrix components \eqref{DPOT} as
\begin{subequations} \label{mmat}
\begin{align}
V_{00}(\vb*{r},\vb*{r}_{0}) \rightarrow V_{Q\overline{Q}}(\vb*{r}) &=
\mel*{\zeta_{Q\overline{Q}}}{H_\text{static}^\text{lf}(\vb*{r})}{\zeta_{Q\overline{Q}}}
\label{qqmat}\\
V_{11}(\vb*{r},\vb*{r}_{0}) \rightarrow V_{M_1\overline{M}_2}(\vb*{r}) &=
\mel*{\zeta_{M_1\overline{M}_2}}{H_\text{static}^\text{lf}(\vb*{r})}{\zeta_{M_1\overline{M}_2}}
\label{mmmat}\\
V_{01}(\vb*{r},\vb*{r}_{0}) \rightarrow V_\text{mix}(\vb*{r}) &=
\mel*{\zeta_{Q\overline{Q}}}{H_\text{static}^\text{lf}(\vb*{r})}{\zeta_{M_1\overline{M}_2}}.
\label{mixmat}
\end{align}
\end{subequations}

Let us realize that having associated each component of the wave function with pure $Q\overline{Q}$ or pure meson-meson,
we can easily incorporate to the kinetic energy operator the fact that the reduced mass of the meson-meson component,
$\mu_{M_1\overline{M}_2}$, is different from $\mu_{Q\overline{Q}}$. Hence, we shall use $- \frac{\hbar^2}{2 \mu_{Q\overline{Q}}}\laplacian$
and $ -\frac{\hbar^2}{2 \mu_{M_1\overline{M}_2}}\laplacian$ for the kinetic energy operators of the $Q\overline{Q}$
and meson-meson components respectively. (Note that this improvement is possible only in the diabatic framework.)

Then, the bound state equations read
\begin{widetext}
\begin{subequations} \label{CD}
\begin{align}
\pqty{-\frac{\hbar^{2}}{2\mu_{Q\overline{Q}}}\laplacian+V_{Q\overline{Q}}(\vb*{r})-E}\psi_{Q\overline{Q}}(\vb*{r})+
V_\text{mix}(\vb*{r})\psi_{M_{1}\overline{M}_{2}}(\vb*{r}) &= 0 \label{C1} \\
\pqty{-\frac{\hbar^{2}}{2\mu_{M_{1}\overline{M}_2}}\laplacian+ V_{M_1\overline{M}_2}(\vb*{r})-E}
\psi_{M_{1}\overline{M}_2}(\vb*{r})+{V_\text{mix}(\vb*{r})}\psi_{Q\overline{Q}}(\vb*{r}) &= 0 ,\label{C2}
\end{align}
\end{subequations}
\end{widetext}
or in matrix notation
\begin{equation}
\pqty{\mathrm{K} + \mathrm{V}(\vb*{r})} \Psi(\vb*{r}) = E \Psi(\vb*{r})
\label{CM}
\end{equation}
where $\mathrm{K}$ is the kinetic energy matrix
\begin{equation}
\mathrm{K} \equiv \pmqty{ -\frac{\hbar^{2}}{2\mu_{Q\overline{Q}}}\laplacian & 0 \\
0 &  -\frac{\hbar^{2}}{2\mu_{M_{1}\overline{M}_2}}\laplacian} ,
\end{equation}
$\mathrm{V}(\vb*{r})$ is the diabatic potential matrix
\begin{equation}
\mathrm{V}(\vb*{r}) \equiv
\pmqty{V_{Q\overline{Q}}(\vb*{r}) & V_\text{mix}(\vb*{r}) \\ {V_\text{mix}(\vb*{r})} &   V_{M_1\overline{M}_2}(\vb*{r})} ,
\label{VM}
\end{equation}
and $\Psi(\vb*{r})$ is a column vector notation for the wave function:
\begin{equation}
\Psi(\vb*{r}) \equiv \pmqty{\psi_{Q\overline{Q}}(\vb*{r}) \\ \psi_{M_{1}\overline{M}_2}(\vb*{r})} .
\end{equation}
In this notation the normalization of the wavefunciton reads
\begin{equation}
\int \dd{\vb*{r}} \Psi^\dagger(\vb*{r})\Psi(\vb*{r}) = \mathcal{P}(Q\overline{Q}) + \mathcal{P}(M_1\overline{M}_2)  = 1
\end{equation}
where we have defined the $Q\overline{Q}$ probability
\begin{equation}
\mathcal{P}(Q\overline{Q}) \equiv \int \dd{\vb*{r}}\abs*{\psi_{Q\overline{Q}}(\vb*{r})}^2
\end{equation}
and the meson-meson probability
\begin{equation}
\mathcal{P}(M_1\overline{M}_2) \equiv  \int \dd{\vb*{r}}\abs*{\psi_{M_{1}\overline{M}_2}(\vb*{r})}^2.
\end{equation}

The multichannel {\schr} equation \eqref{CD}, or equivalently \eqref{CM}, defines formally the diabatic approach for
the description of the heavy-quark meson system.

\subsection{\label{mixpotsec}Mixing potential}

To solve \eqref{CD} we need to know the diabatic potential matrix Eq.~\eqref{VM}. Regarding the diagonal element
$V_{Q\overline{Q}}(\vb*{r})$, we see from
\eqref{qqmat} that it corresponds to the expectation value of the
static energy operator in the light field state
associated with a pure $Q\overline{Q}$ configuration.
This can be identified with the ground state static energy
calculated in quenched lattice QCD, see Fig.~\ref{cross},
given by the Cornell potential
\begin{equation}
V_{Q\overline{Q}}(\vb*{r}) = V_\text{C}(r).
\label{VQQDEF}
\end{equation}
In the same way, from \eqref{mmmat} we identify the other diagonal term $V_{M_1\overline{M}_2}(\vb*{r})$ with the static energy
associated with a pure meson-meson configuration, given by
the threshold mass $T_{M_1\overline{M}_2}$ (the sum of the meson masses)
\begin{equation}
V_{M_1\overline{M}_2}(\vb*{r}) = T_{M_1\overline{M}_2} \equiv m_{M_1} + m_{\overline{M}_2} ,
\label{VMMDEF}
\end{equation}
up to one pion exchange effects that we do not consider here.

As for the off-diagonal term, the mixing potential $V_\text{mix}(\vb*{r})$, we can use the eigenvalues of the diabatic
potential matrix to derive its form. As shown in Appendix~\ref{apdxADT}, these eigenvalues correspond to the static
energy levels that are calculated in unquenched lattice QCD which have been pictorially represented
in Fig.~\ref{cross}. More precisely, the eigenvalues of the diabatic potential matrix are the
two solutions $V_\pm(\vb*{r})$ of the secular equation
\begin{equation}
\det{\mathrm{V}(\vb*{r}) - V_\pm(\vb*{r}) \mathbb{I}} = 0
\end{equation}
where $\mathbb{I}$ is the identity matrix. These solutions read
\begin{equation}
\begin{split}
V_\pm(\vb*{r}) = &\frac{V_\text{C}(r) + T_{M_1\overline{M}_2}}{2} \\ 
&\pm \sqrt{\pqty{\frac{V_\text{C}(r) -
T_{M_1\overline{M}_2}}{2}}^2 + V_\text{mix}(\vb*{r})^2},
\end{split}
\end{equation}
from which we obtain
\begin{equation}
\abs{V_\text{mix}(r)} = \frac{\sqrt{\pqty{V_+(r) - V_-(r)}^2 - \pqty{V_\text{C}(r) - T_{M_1\overline{M}_2}}^2}}{2},
\label{VMIX}
\end{equation}
where we have dropped the vector notation for $\vb*{r}$ as the energy levels calculated in
lattice QCD depend only on the modulus $r=\abs{\vb*{r}}$.

Eq.~\eqref{VMIX} tells us that a detailed calculation of the mixing potential $\abs{V_\text{mix}(r)}$ from \textit{ab initio} lattice
data on $V_\pm(r)$ is possible. As a matter of fact, an effective parametrization of $V_\text{mix}(r)$ from lattice data has been proposed
\cite{Bul19, Bic20}. While we encourage more work along this direction, we resort to general arguments
to get the shape of $\abs{V_\text{mix}(r)}$. In this regard, the general form of the curves $V_{+}(r)$ and $V_{-}(r)$
near any threshold, reflecting the physical picture of the
$Q\overline{Q}$ -- meson-meson mixing, is expected to be similar as it happens to be the case when two thresholds
are incorporated into the lattice
calculation \cite{Bal05,Bul19}. Furthermore, the same form is expected for $Q=b$ and $Q=c$ since the underlying
mixing mechanism (string breaking) is the same.
Therefore, we shall proceed to a parametrization of $\abs{V_\text{mix}(r)}$ according to this general form, and we shall
rely on phenomenology to fix the values of the parameters.

Let us begin by observing that unquenched lattice QCD results show that
\begin{equation}
\abs{V_+(r) - V_-(r)} \ge \abs{V_C (r) - T_{M_{1} \overline{M}_2}}
\end{equation}
for every value $r$, and that at the crossing radius $r_\text{c}^{M_{1} \overline{M}_2}$, defined by
\begin{equation}
V_\text{C}\bigl(r_\text{c}^{M_{1} \overline{M}_2}\bigr)=T_{M_{1} \overline{M}_2},
\end{equation}
$\abs{V_\text{mix}(r)}$ gets approximately its maximum value
\begin{equation}
\max_r \abs{V_\text{mix}(r)} \approx \abs{V_\text{mix}\bigl(r_\text{c}^{M_{1} \overline{M}_2}\bigr)} = \frac{\Delta}{2},
\end{equation}
with $\Delta$ being the distance of the static energy levels at the crossing
radius
\begin{equation}
\Delta \equiv \abs{V_+\bigl(r_\text{c}^{M_{1} \overline{M}_2}\bigr) - V_-\bigl(r_\text{c}^{M_{1} \overline{M}_2}\bigr)}.
\end{equation}
On the other hand we have
\begin{equation}
V_{-}(r)\approx V_\text{C}(r) \qand V_{+}(r)\approx T_{M_{1} \overline{M}_2}
\end{equation}
for $r\ll r_\text{c}^{M_{1} \overline{M}_2}$, and
\begin{equation}
V_{-}(r)\approx T_{M_{1} \overline{M}_2} \qand V_{+}(r)\approx V_\text{C}(r)
\end{equation}
for $r\gg r_\text{c}^{M_{1} \overline{M}_2}$, so that
\begin{equation}
(V_{+}(r) - V_{-}(r))^{2} \approx (V_\text{C}(r) - T_{M_{1} \overline{M}_2})^{2}
\end{equation}
far from the crossing radius $r_\text{c}^{M_{1} \overline{M}_2}$.
Consequently, from \eqref{VMIX} we obtain that $V_\text{mix}(r)$ vanishes in both asymptotic limits:
\begin{equation}
\lim_{r \to 0}V_\text{mix}(r) = \lim_{r \to \infty}V_\text{mix}(r) = 0 .
\end{equation}

To summarize, lattice QCD indicates that the mixing potential $\abs{V_\text{mix}(r)}$ approaches a maximum value of $\Delta/2$ at
$r\approx r_\text{c}^{M_{1} \overline{M}_2}$ and vanishes asymptotically as the distance from the crossing
radius increases. The simplest
parametrization that takes into account these behaviors, thus providing a
good fit to lattice QCD calculations of $V_{\pm}(r)$, is a Gaussian shape:
\begin{equation}
\abs{V_\text{mix}(r)} =\frac{\Delta}{2}\exp{-\frac{\pqty{V_\text{C}(r)-T_{M_{1} \overline{M}_2}}^{2}}{2\Lambda^{2}}}
\label{VMIXDEF}
\end{equation}
where $\Lambda$ is a parameter with dimensions of energy. To better understand the physical meaning of
$\Lambda$ we write it in terms of the string tension $\sigma$ as
\begin{equation}
\Lambda\equiv\sigma\rho
\end{equation}
where $\rho$ has now dimensions of length. Then at distances for which
$V_\text{C}(r)\approx\sigma r + m_Q + m_{\overline{Q}} - \beta$ the mixing potential can be also written as
\begin{equation}
\abs{V_\text{mix}(r)} \approx\frac{\Delta}{2}\exp{-\frac{\bigl(r-r_\text{c}^{M_{1} \overline{M}_2}\bigr)^{2}}{2\rho^{2}}}
\tag{\ref*{VMIXDEF}$^\prime$}
\end{equation}
from which it is clear that $\rho$, the width of the Gaussian curve, fixes a radial scale for the mixing.

\subsection{\label{angsec}Configuration mixing}

The knowledge of the diabatic potential matrix is quite equivalent to the knowledge of the $r$-dependent change of basis matrix
from $\Bqty{\ket{\zeta_0(r)}, \ket{\zeta_1(r)}}$ to $\Bqty{\ket{\zeta_0(r_0)},\ket{\zeta_1(r_0)}}$.
Let us name, according to our previous notation,
$\ket*{\zeta_-(r)}\equiv\ket{\zeta_0(r)}$ and $\ket*{\zeta_+(r)}\equiv\ket{\zeta_1(r)}$ the ground and excited
states of the light fields, with static energies $V_-(r)$ and $V_+(r)$ respectively.
These are related to the $Q\overline{Q}$ and meson-meson states $\ket*{\zeta_{Q\overline{Q}}}\equiv \ket{\zeta_0(r_0)}$
and $\ket*{\zeta_{M_1\overline{M}_2}}\equiv \ket{\zeta_1(r_0)}$ via
\begin{subequations} \label{mixangdef}
\begin{align}
\ket{\zeta_-(r)} &= \cos(\theta(r)) \ket*{\zeta_{Q\overline{Q}}} + \sin(\theta(r)) \ket*{\zeta_{M_1\overline{M}_2}} \\
\ket{\zeta_+(r)} &= \cos(\theta(r)) \ket*{\zeta_{M_1\overline{M}_2}} - \sin(\theta(r)) \ket*{\zeta_{Q\overline{Q}}}
\end{align}
\end{subequations}
where $\theta(r)$ is the \emph{mixing angle} between the $Q\overline{Q}$ and meson-meson configurations.

As explained in Appendix~\ref{apdxADT}, the change of basis matrix connecting the two
sets of states,
\begin{equation}
\pmqty{\ket{\zeta_-(r)} \\ \ket{\zeta_+(r)}} = \mathrm{A}^\dagger(r) \pmqty{\ket*{\zeta_{Q\overline{Q}}} \\ \ket*{\zeta_{M_1\overline{M}_2}}}
\end{equation}
with
\begin{equation}
\mathrm{A}(r) \equiv \pmqty{\cos(\theta(r)) & -\sin(\theta(r)) \\ \sin(\theta(r)) & \cos(\theta(r))},
\end{equation}
is also the matrix that diagonalizes the diabatic potential matrix. Therefore it is possible to extract the mixing angle $\theta$ from the matrix equation
\begin{equation}
\mathrm{A} (r) \mathrm{V}(r) \mathrm{A}^\dagger (r) = \text{diag}(V_-(r), V_+(r))
\label{mateqq}
\end{equation}
where $\text{diag}(V_-(r), V_+(r))$ is a diagonal 2$\times$2 matrix containing the unquenched static light field energies.
It is sufficient to take any off-diagonal element of Eq.~\eqref{mateqq} to obtain
\begin{equation}
V_\text{mix}(r) \cos (2\theta(r)) = \frac{T_{M_1 \overline{M}_2} - V_\text{C}(r)}{2} \sin(2\theta(r))
\end{equation}
from which we get the mixing angle as
\begin{equation}
\theta (r) = \frac{1}{2} \arctan\biggl(\frac{2 V_\text{mix}(r)}{T_{M_1 \overline{M}_2} - V_\text{C}(r)}\biggr).
\label{mixangeq}
\end{equation}
Furthermore, from this expression of the mixing angle and from Eqs.~\eqref{mixangdef} we can also calculate the NACTs:
\begin{subequations}
\begin{align}
\vb*{\tau}_{00}(r) &= \vb*{\tau}_{11}(r) = 0 \\
\vb*{\tau}_{01}(r) &= - \vb*{\tau}_{10}(r)
\end{align}
\end{subequations}
with
\begin{equation}
\vb*{\tau}_{0 1}(r) \equiv \braket{\zeta_-(r)}{\grad\zeta_+(r)} = (\mathrm{A} (r) \grad \mathrm{A}^\dagger (r))_{0 1} = \vu*{r} \dv{\theta}{r} .
\end{equation}
Therefore the NACTs only vanish for values of $r$ where $\theta$
is constant. This happens for small (big) values of $r$ where $\theta$ is $0$ $(\pi/2)$,
corresponding to no mixing between the $Q\overline{Q}$ and meson-meson configurations
in the light field eigenstates.

\subsection{General case}

The multichannel {\schr} equation \eqref{CD} defines the heavy quark meson system when only one threshold is considered, but in general
it may be necessary to incorporate several meson-meson thresholds.
In such a case one has to extend the formalism, what is is more easily done in the matrix notation \eqref{CM}.

The generalization of the kinetic energy matrix is straightforward:
\begin{equation}
\mathrm{K} =
\begin{pmatrix}
 -\frac{\hbar^{2}}{2\mu_{Q\overline{Q}}}\laplacian 	& 												& 		&
											\\
										& -\frac{\hbar^{2}}{2\mu^{(1)}_{M\overline{M}}}\laplacian 	& 		&
											\\
										& 												& \ddots 	&
											\\
										& 												& 		&
-\frac{\hbar^{2}}{2\mu^{(N)}_{M\overline{M}}}\laplacian 	\\
\end{pmatrix}
\end{equation}
where $\mu_{M\overline{M}}^{(i)}$ with $i=1,\dots,N$ is the reduced mass of the $i$-th meson-meson component,
$N$ is the number of meson-meson thresholds, and matrix elements equal to zero are not displayed.

As for the extension of the diabatic potential matrix \eqref{VM}, the presence of interaction
terms between different meson-meson components  would make not practicable our
procedure to extract the mixing potentials.
Following what it is usually done in molecular physics \cite{Bae06}, we neglect some interactions between components.
Namely, in line with lattice QCD studies of string breaking \cite{Bul19}, we assume that different
meson-meson components do not interact with each other.

It seems reasonable to think that this is a good approximation when dealing with relatively
narrow, well-separated thresholds. If so, we may consider the uncertainty of this approximation
to be proportional to the ratio between the average of the threshold widths and the threshold
mass difference. More precisely, for values of this ratio smaller than one we expect the
threshold-threshold interaction to be negligible. According to this, we restrict our study
to non-overlapping, narrow thresholds.

Then, the diabatic potential matrix with $N$ thresholds reads
\begin{equation}
\mathrm{V}(r) =
\begin{pmatrix}
V_\text{C}(r)				& V_\text{mix}^{(1)}(r) 		& \hdots 	& V_\text{mix}^{(N)}(r) 	\\
{ V_\text{mix}^{(1)}(r)}		& T_{M\overline{M}}^{(1)}	&		&					\\
\vdots 					&						& \ddots	&					\\
{V_\text{mix}^{(N)}(r)}		&						&		& T_{M\overline{M}}^{(N)}\\
\end{pmatrix}
\label{dpotmany}
\end{equation}
where $V_\text{C}(r)$ stands for the Cornell potential, $T_{M\overline{M}}^{(i)}$ for the mass
of the $i$-th threshold and $V_\text{mix}^{(i)}(r)$
for the mixing potential between the $Q\overline{Q}$ and the $i$-th meson-meson components.

\begin{figure}
\includegraphics{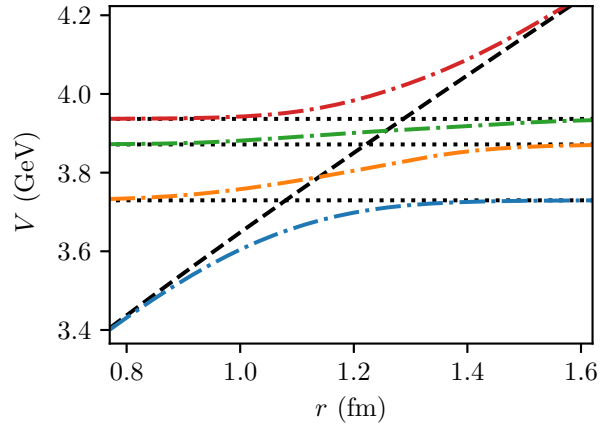}
\caption{Static energies. Dashed line: $c\overline{c}$ (Cornell) potential~\eqref{CPOT} with $\sigma=925.6$~MeV/fm,
$\chi=102.6$~MeV~fm, $\beta = 855$~MeV and $m_c=1840$~MeV.
Dotted lines: meson-meson thresholds ($D \overline{D}$, $D \overline{D}^*$, $D_s \overline{D}_s$).
Dash-dotted lines: $r$-dependent eigenvalues of the diabatic potential matrix. For the sake of simplicity we have assumed the same mixing potential
parameters for all the meson-meson components: $\Delta_{c\overline{c}} = 130$~MeV and $\rho_{c\overline{c}} = 0.3$~fm.}
\label{many}
\end{figure}

In Fig.~\ref{many} we draw the eigenvalues of this matrix for $c\overline{c}$ and the first three open flavor meson-meson thresholds.

The diabatic potential matrix \eqref{dpotmany} can be regarded as a generalization of the two threshold model
of string breaking introduced in \cite{Bul19}, the two main differences being that in our study each dynamical quark flavor
can introduce more than one threshold and that we have parametrized the coupling between quark-antiquark and meson-meson
components with a Gaussian instead of a constant.

Let us add that even tough there is presumably an infinite number of possible meson-meson components,
in practice one needs to consider only a limited subset of them when searching for bound
states. As a matter of fact, a meson-meson component hardly plays any role in the
composition of a bound state whose mass lies far below the corresponding threshold.

\subsection{Quantum numbers}

Heavy-quark meson states are characterized
by quantum numbers $I^{G}\pqty{J^{PC}}$ where $I$, $G$, $J$, $P$, $C$ stand for the isospin, G-parity, total
angular momentum, parity, and charge conjugation quantum numbers respectively.

Let us focus on isoscalars $I=0$ heavy-quark mesons, for which $G=C$. Since the diabatic potential matrix is spherically symmetric
and spin-independent, the $Q\overline{Q}$ component of the wave function can be characterized by the relative orbital angular momentum
quantum number $l_{Q\overline{Q}}$, the total spin $s_{Q\overline{Q}}$, the total angular momentum $J$ and its projection $m_J$ so that
\begin{subequations}\label{quanums}
\begin{align}
\vb*{L}^2_{Q\overline{Q}} Y_{l_{Q\overline{Q}}}^{m_l}(\vu*{r}) &= \hbar^2 l_{Q\overline{Q}} (l_{Q\overline{Q}} + 1)
Y_{l_{Q\overline{Q}}}^{m_l}(\vu*{r}) \\
\vb*{S}^2_{Q\overline{Q}} \xi_{s_{Q\overline{Q}}}^{m_s} &= \hbar^2 s_{Q\overline{Q}} (s_{Q\overline{Q}} + 1) \xi_{s_{Q\overline{Q}}}^{m_s} \\
\vb*{J}^2 \bqty{Y_{l_{Q\overline{Q}}}(\vu*{r}) \xi_{s_{Q\overline{Q}}}}_J^{m_J} &=
\hbar^2 J (J+1)  \bqty{Y_{l_{Q\overline{Q}}}(\vu*{r}) \xi_{s_{Q\overline{Q}}}}_J^{m_J} \\
J_z \bqty{Y_{l_{Q\overline{Q}}}(\vu*{r}) \xi_{s_{Q\overline{Q}}}}_J^{m_J} &=
\hbar \, m_J \bqty{Y_{l_{Q\overline{Q}}}(\vu*{r}) \xi_{s_{Q\overline{Q}}}}_J^{m_J}
\end{align}
\end{subequations}
where $Y_{l}^{m_l}(\vu*{r})$ is the spherical harmonic of degree $l$, $\xi_{s}^{m_s}$ is the eigenstate of the total $Q\overline{Q}$ spin
and $\bqty{Y_{l_{Q\overline{Q}}}(\vu*{r}) \xi_{s_{Q\overline{Q}}}}_J^{m_J}$ is a shorthand notation for the sum
\begin{equation}
\bqty{Y_{l_{Q\overline{Q}}} (\vu*{r}) \xi_{s_{Q\overline{Q}}}}_J^{m_J} \equiv
\sum_{m_l, m_s} C_{l_{Q\overline{Q}}, s_{Q\overline{Q}}, J}^{m_l, m_s, m_J} Y_{l_{Q\overline{Q}}}^{m_l} (\vu*{r}) \xi_{s_{Q\overline{Q}}}^{m_s}
\end{equation}
where $C_{l, s, J}^{m_l, m_s, m_J}$ is the Clebsch-Gordan coefficient. Given this set of quantum numbers,
the $Q\overline{Q}$ component of the wave function can be factorized as
\begin{equation}
\psi_{Q\overline{Q}}(\vb*{r}) = u^{(Q\overline{Q})}_{E, l_{Q\overline{Q}}}(r) \bqty{Y_{l_{Q\overline{Q}}}(\vu*{r}) \xi_{s_{Q\overline{Q}}}}_J^{m_J}
\end{equation}
where $u^{(Q\overline{Q})}_{E, l_{Q\overline{Q}}}(r)$ is the $Q\overline{Q}$ radial wave function.

The same can be done for the meson-meson components of the wave function, considering the meson-meson relative orbital angular momentum
$l_{M_1\overline{M}_2}$ and the sum of their spins $s_{M_1\overline{M}_2}$. Therefore, with a straightforward extension of the above notation we write
\begin{equation}
\psi_{M_1\overline{M}_2}(\vb*{r}) =  u^{(M_1\overline{M}_2)}_{E, l_{M_1\overline{M}_2}}(r)
\bqty{Y_{l_{M_1\overline{M}_2}}(\vu*{r}) \xi_{s_{M_1\overline{M}_2}}}_J^{m_J}.
\end{equation}
Note that for the spectroscopic state to have a definite value of $J$, the $Q\overline{Q}$ and all the meson-meson
components must have the same total angular momentum, hence the unified notation for $J$.

A bound state made of $Q\overline{Q}$ and meson-meson has definite parity and $C$\nobreakdash-parity only if all the wave function components
have the same parity under these transformations. This requirement translates into different conditions depending on whether
the wave function component is associated with $Q\overline{Q}$ or meson-meson. For the
$Q\overline{Q}$ component, $P$ and $C$ quantum numbers are given by
\begin{equation}
P = (-1)^{l_{Q\overline{Q}} + 1} \qand C = (-1)^{l_{Q\overline{Q}} + s_{Q\overline{Q}}}.
\end{equation}
On the other hand, for each meson-meson component one has
\begin{equation}
P = P_{M_1} P_{\overline{M}_2} (-1)^{l_{M_1\overline{M}_2}}
\end{equation}
where $P_{M}$ is the parity of the meson. As for $C$\nobreakdash-parity, one has to consider two distinct cases: if $M_1 = M_2$
the $C$\nobreakdash-parity of the meson-meson component is given by
\begin{equation}
C = (-1)^{l_{M_1 \overline{M}_2} + s_{M_1 \overline{M}_2}},
\label{Cmesmes}
\end{equation}
if otherwise $M_1 \ne M_2$ one can build both positive and negative $C$\nobreakdash-parity states
\begin{equation}
C \ket*{M_1 \overline{M}_2}_\pm = \pm \ket*{M_1 \overline{M}_2}_\pm
\end{equation}
taking the linear combinations
\begin{equation}
\ket*{M_1 \overline{M}_2}_\pm \equiv \frac{1}{\sqrt{2}}\pqty{\ket*{M_1 \overline{M_2}}_0 \pm
\mathcal{C}_{M_1 \overline{M}_2} \ket*{M_2 \overline{M}_1}_0}
\label{Cpm}
\end{equation}
with $\ket*{M_1 \overline{M}_2}_0$ being the isospin singlet state obtained from the combination of the $M_1$ and $\overline{M}_2$ isomultiplets and
\begin{equation}
\mathcal{C}_{M_1 \overline{M}_2}\equiv (-1)^{l_{M_1 \overline{M}_2} + s_{M_1 \overline{M}_2}
+ l_{M_1} + l_{\overline{M}_2} + s_{M_1} + s_{\overline{M}_2} + j_{M_1} + j_{\overline{M}_2}}
\label{Cm1m2}
\end{equation}
where $l_M$ is the internal orbital angular momentum of the meson, $s_M$ its internal spin and $j_M$ its total spin.
The derivation of Eqs.~\eqref{Cpm} and \eqref{Cm1m2} is detailed in Appendix~\ref{apdxcpar}.

\subsection{Bound state solutions}

Given a spherically-symmetric and spin-independent diabatic potential matrix, each $Q\overline{Q}$ configuration with a distinct value of
$(l_{Q\overline{Q}},s_{Q\overline{Q}})$ can be treated as a channel \textit{per se}, and the same can be said for each meson-meson configuration
with a distinct value of $(l_{M_1\overline{M}_2},s_{M_1\overline{M}_2})$. Then finding the spectrum of a given $J^{PC}$ family
boils down to solving a multichannel, spherical {\schr} equation involving only those channels with the corresponding $J^{PC}$ quantum numbers.

One should realize though that a complete numerical nonperturbative solution of the spectroscopic equations \eqref{CM}
is only possible for energies below the lowest $J^{PC}$ threshold. Above it the asymptotic behavior of its meson-meson
component as a free wave, against the confined $Q\overline{Q}$ wave, prevents obtaining a physical solution. Nonetheless,
an approximate physical solution for energies above threshold is still possible, under the assumption that the effect of an open
threshold on the above-lying bound states can be treated perturbatively. More in detail, we proceed in the following way:
\begin{enumerate}[i)]

\item We build the effective $J^{PC}$ diabatic potential matrix out of the Cornell $Q\overline{Q}$ potential, the threshold masses, and the
$Q\overline{Q}$ -- meson-meson mixing potentials.

\item \label{it1} We solve the spectroscopic equations for energies up to the lowest $J^{PC}$ threshold mass, and we analyze the
$(n \,{^{2S+1}\!L}_{J})$ $Q\overline{Q}$ and meson-meson content of the bound states.

\item \label{it2}We build a new $J^{PC}$ diabatic potential matrix neglecting the $Q\overline{Q}$ coupling to the lowest (first) threshold. We solve it for
energies in between the lowest and the second thresholds and discard as spurious any solution containing a $(n \,{^{2S+1}\!L}_{J})$
$Q\overline{Q}$ state entering in the bound states calculated in \ref{it1}). The rationale underlying this step is that a given spectral state in between the
lowest and the second thresholds containing such a $(n \,{^{2S+1}\!L}_{J})$ $Q\overline{Q}$ component would become, when the lowest
threshold were incorporated, the bound state below threshold containing it found in \ref{it1}).

\item We build a new $J^{PC}$ diabatic potential matrix by neglecting the coupling to the lowest threshold and to the second one. We solve it for
energies in between the second and the third thresholds and discard as spurious any solution containing a $(n \,{^{2S+1}\!L}_{J})$
$Q\overline{Q}$ state entering in the bound states calculated in \ref{it1}) and \ref{it2}), and so on.

\item We assume that corrections to the physical states thus obtained due to the coupling with open thresholds can be implemented perturbatively.

\end{enumerate}

The formulation of an appropriate perturbative scheme for the calculation of these corrections,
giving rise to mass shifts as well as to decay
widths to open flavor meson-meson states, will be the subject of a forthcoming paper. 
On the other hand there are certainly more corrections to the spectrum that are not
included in our treatment, in particular those due to spin interactions.
Regarding the $Q\overline{Q}$ component, these effects can be
incorporated by adding spin-dependent operators (e.g.\ spin-spin, spin-orbit, tensor) to the Cornell
potential, what has proven to be very effective for a detailed description
of the low-lying spectral states \cite{GI85}. As for meson-meson components, the part of these
corrections involving quark and antiquark within the same heavy-light meson are included through the meson
masses, whereas the remaining ones can be implemented through the one pion exchange interaction
between mesons.

Assuming that these additional energy contributions (fine and hyperfine splittings,
one pion exchange corrections, mass shifts from coupling to open thresholds) can be
taken into account using perturbation theory,  we shall concentrate henceforth on the calculation
of the ``unperturbed'' heavy-quark meson spectrum. The technical procedure followed to solve
the spectroscopic equations is detailed in
Appendices~\ref{apdxvar} and \ref{apdxlag}.

\section{\label{sec5}Charmonium-like mesons}

The formalism we have developed in the previous sections can be tested in
charmonium-like mesons (heavy mesons containing $c\overline{c}$) where, unlike in the bottomonium-like case, there are
several well-established experimental candidates
for unconventional isoscalar states, presumably containing significant
meson-meson components. In particular, we center on
isoscalar states with masses up to about $4.1$~GeV, for which the relevant
thresholds have very small widths and do not overlap. A list of these
thresholds is shown in Table~\ref{tablist}.

\begin{table}
\begin{ruledtabular}
\begin{tabular}{cc}
$M_{1}\overline{M}_{2}$		& $T_{M_{1}\overline{M}_{2}}$ (MeV)	\\
\hline
$D\overline{D}$ 			& $3730$							\\
$D\overline{D}^*(2007)$ 		& $3872$ 							\\
$D_{s}^{+}D_{s}^{-}$			& $3937$							\\
$D^*(2007)\overline{D}^*(2007)$	& $4014$							\\
$D_{s}^{+}D_{s}^{*-}$ 		& $4080$							\\
\end{tabular}
\end{ruledtabular}
\caption{\label{tablist}Low-lying open charm meson-meson thresholds $M_{1}\overline{M}_{2}$.
Threshold masses $T_{M_{1}\overline{M}_{2}}$ from the charmed and charmed
strange meson masses quoted in \cite{PDG20}.}
\end{table}

The possible values of the meson-meson relative orbital angular momentum contributing
to any given set of quantum numbers $J^{PC}$ are shown in Table~\ref{quanum}. Note that
we use the common notation $D_{(s)}$ to refer to charmed as well as to charmed strange
mesons and the shorthand notation $D_{(s)}\overline{D}^*_{(s)}$ for
the meson-meson $C$\nobreakdash-parity eigenstate defined by Eq.~\eqref{Cpm}.

\begin{table}
\begin{ruledtabular}
\begin{tabular}{cccc}
$J^{PC}$	& $l_{D_{(s)} \overline{D}_{(s)}}$	& $l_{D_{(s)} \overline{D}_{(s)}^*}$	& $l_{D_{(s)}^{*} \overline{D}_{(s)}^*}$ 	\\
\hline
$0^{++}$	& $0$						& 								& 0,\,2							\\
$1^{++}$	& 							& $0,\,2$							& $2$							\\
$2^{++}$	& $2$						& $2$							& $0,\,2$ 							\\
$1^{--}$	& $1$						& $1$							& $1,\,3$							\\
\end{tabular}
\end{ruledtabular}
\caption{\label{quanum}Values of $l_{M_1\overline{M}_2}$ corresponding to meson-meson configurations
with definite values of $J^{PC}$. A missing entry means that the particular meson-meson configuration cannot
form a state with the corresponding quantum numbers.}
\end{table}

In order to calculate the heavy-quark meson bound states we have to fix the
values of the parameters. For the Cornell potential \eqref{CPOT} we use the standard values \cite{Eic94}%
\begin{subequations} \label{params}
\begin{align}
\sigma &= 925.6 \text{~MeV/fm}, \\
\chi &= 102.6 \text{~MeV~fm}, \\
m_{c} &= 1840\text{~MeV}
\end{align}
and we choose
\begin{equation}
\beta = 855\text{~MeV}
\end{equation}
\end{subequations}
in order to fit the $2s$ center of gravity. Let us note that
one could alternatively choose to fit the $1s$ or $1p$ centers of
gravity, or to get a reasonable fit to the three of them. Our choice is based on the
assumption that relativistic mass effects in the higher states, which are at
least in part incorporated in $\beta$, are expected to deviate
less from those in the $2s$ states.

We should also mention that the value of the charm quark mass we use
is completely consistent with the one needed to
correctly describe $c\overline{c}$ electromagnetic decays within the Cornell
potential model framework \cite{Bru20}.

The low-lying spectrum from this Cornell potential for $J^{++}$ and
$1^{--}$ isoscalar states is shown in Table~\ref{cortable}.

\begin{table}
\begin{ruledtabular}
\begin{tabular}{cccc}
$J^{PC}$ & $nl$ & $M_{c\overline{c}}$ (MeV) & $M_\text{cog}^\text{Expt}$ (MeV)\\
\hline
\multirow{4}*{$1^{--}$} & $1s$ & $3082.5$ & $3068.65\pm0.13$\\
& $2s$ & $3673.2$ & $3674.0\pm0.3$\\
& $1d$ & $3795.8$ & \\
& $3s$ & $4097.0$ & \\
\\
\multirow{2}*{$(0,1,2)^{++}$} & $1p$ & $3510.9$ & $3525.30\pm0.11$\\
& $2p$ & $3953.7$ & \\
\end{tabular}
\end{ruledtabular}
\caption{\label{cortable}Calculated $J^{++}$ and $1^{--}$ charmonium masses, $M_{c\overline
{c}}$, for spectroscopic $nl$ states from the Cornell potential \eqref{CPOT}
with parameters \eqref{params}. Experimental mass
centroids from \cite{PDG20}, $M_\text{cog}^\text{Expt}$, are listed for
comparison.}
\end{table}

For the lowest $J^{++}$ states it is worth to remark, apart from the good average mass description, the
excellent fit to the mass of the lowest $1^{++}$ state, $\chi_{c_{1}}(1p)$
($3510.9$~MeV versus the experimental mass $3510.7$
MeV). However, an accurate fit of the lowest $(0,2)^{++}$ masses, in particular for
$\chi_{c_{0}}(1p)$, would require the incorporation of
correction terms (e.g.\ spin-spin, spin-orbit, tensor) to the Cornell radial potential. As
for the first excited $J^{++}$ states one could expect a similar situation (the $2s$ states
lie in between the $1p$ and $2p$ ones$)$ in the absence of threshold effects
that we analyse in what follows.

As for the parameters of the mixing potential \eqref{VMIXDEF},
we have to rely on phenomenology since the only lattice information available is
for $b\overline{b}$. We fix them by requiring that our diabatic treatment
fits the mass of some unconventional experimental state lying close below
threshold. In particular, we can use the mass of $\chi_{c1}(3872)$, a
well-established experimental resonance lying just below the $D\overline
{D}^*$ threshold, to infer the possible of values for $\Delta_{c\overline
{c}}$ and $\rho_{c\overline{c}}$.

As the crossing of the Cornell potential with the $D\overline{{D}}^*$
threshold takes place around $r_\text{c}^{D\overline
{D}^*}=1.76$~fm, we conservatively vary
$\rho_{c\overline{c}}$ from $0.1$~fm to $0.8$~fm, this last value
corresponding to almost half of $r_\text{c}^{D\overline{D}^*}$. Then, for
every value of $\rho_{c\overline{c}}$ we get the minimal value of
$\Delta_{c\overline{c}}$ to accurately fit the mass of $\chi_{c1}(3872)$. The calculated values are
listed in Table~\ref{rodel}.

\begin{table}
\begin{ruledtabular}
\begin{tabular}{cc}
$\rho_{c\overline{c}}$ (fm)	& $\Delta_{c\overline{c}}$ (MeV)	\\
\hline
0.1				& 290				\\
0.2				& 165				\\
0.3				& 130				\\
0.4				& 115				\\
0.5				& 108				\\
0.6				& 104				\\
0.7				& 102				\\
0.8				& 101				\\
\end{tabular}
\end{ruledtabular}
\caption{\label{rodel}Correlated values of the mixing potential parameters giving rise
to a $0^+(1^{++})$  bound state with a mass close below the $D \overline{D}^*$ threshold.}
\end{table}

\begin{figure}
\includegraphics{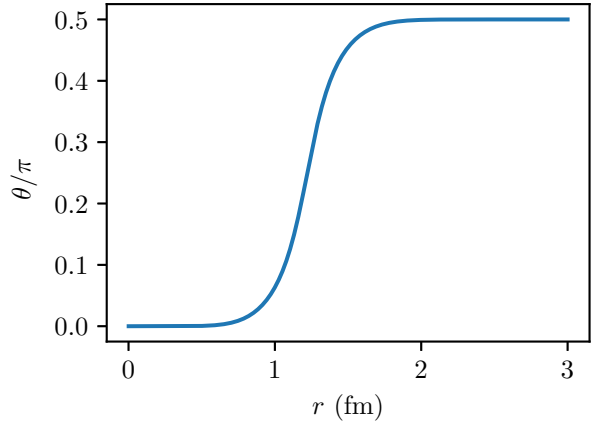}
\caption{\label{mixang}Mixing angle between $c\overline{c}$ and $D \overline{D}^*$ with $\Delta_{c\overline{c}}$=130~MeV,
$\rho_{c\overline{c}}$ = 0.3 fm and Cornell potential parameters \eqref{params}.}
\end{figure}

It should be pointed out that large values of $\Delta_{c\overline{c}}$ would
deform the shape of the avoided energy crossings as compared to the one
calculated in lattice for $b\overline{b}$, against our $b\overline
{b}$ -- $c\overline{c}$ universality arguments for the shape of the mixing potential. On the
other hand, large values of $\rho_{c\overline{c}}$ would make the
mixing angle between the $c\overline{c}$ and a single $M_1 \overline{M}_2$ threshold,
calculated from Eq.~\eqref{mixangeq},
to have an asymptotic behavior in conflict with
the one observed in the lattice under the
natural assumption that this behavior is similar for $b\overline{b}$ and
$c\overline{c}$.
More precisely, unquenched lattice QCD
calculations of the mixing angle \cite{Bal05} show that $\theta$ approaches
$\pi/2$ quite rapidly for $r>r_\text{c}^{M_1\overline{M}_2}$, thus ruling out
a large radial scale for the mixing. Henceforth we use%
\begin{subequations}\label{mixparam}
\begin{equation}
\rho_{c\overline{c}}=0.3\text{ fm}
\end{equation}
for this value gives the most accurate asymptotic behavior of the mixing
angle, see Fig.~\ref{mixang}, and consequently
\begin{equation}
\Delta_{c\overline{c}}=130\text{~MeV}.
\end{equation}
\end{subequations}

\begin{figure}
\includegraphics{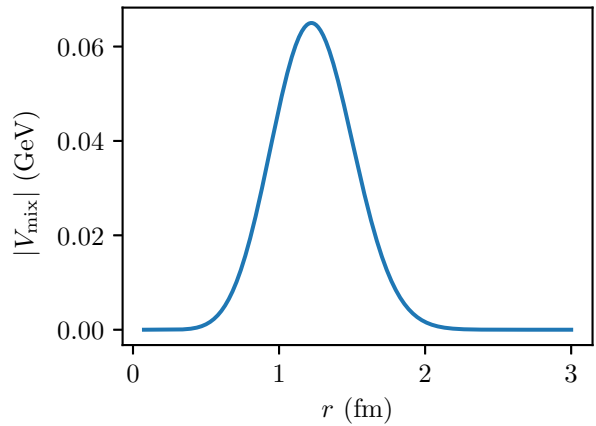}
\caption{\label{mixpot}Mixing potential for $c\overline{c}$ and $D\overline{D}^*$ with $\Delta_{c\overline{c}}$=130~MeV and
$\rho_{c\overline{c}}$ = 0.3 fm.}
\end{figure}

The resulting mixing potential is drawn in Fig.~\ref{mixpot} for $M_1\overline{M}_2=D\overline{D}^*$. For any
other threshold the only difference comes from the substitution of
the threshold mass.

Notice that we have drawn $\abs{V_\text{mix}(r)}$ with no sign prescription for $V_\text{mix}(r)$.
This sign can be reabsorbed as a relative phase between the charmonium and meson-meson components.
For the calculations in this paper a positive sign has been taken. We have checked that
for the observables considered in this article the same results are obtained with a negative sign.
It should be realized though that this could not be the case for other observables.

The calculated spectrum of $J^{++}$ states, containing one $c\overline{c}$
state with $l_{c\overline{c}}=1$ ($1p$ or $2p$), is shown in Table~\ref{charm_jpp_table}.

\begin{table*}
\begin{ruledtabular}
\begin{tabular}{ccdddddd}
$J^{PC}$	& Mass (MeV)	& c\overline{c}	& D \overline{D}	& D \overline{D}^*
	& D_s \overline{D}_s	& D^* \overline{D}^*	& D_s \overline{D}_s^*	\\
\hline
\multirow{2}*{$1^{++}$}	& 3510.0			& 100 \%	& 		& 		& 		& 			& 			\\
		& 3871.7			& 1 \%	& 		&  99 \%	& 		& 			& 			\\
\\
\multirow{2}*{$0^{++}$}	& 3509.1			& 100 \%	& 		& 		& 		& 			& 			\\
		& 3920.4			& 59 \%	& 		& 		& 37 \%	& 4 \%		& 			\\
\\
\multirow{2}*{$2^{++}$}	& 3509.6			& 100 \%	& 		& 		& 		& 			& 			\\
		& 3933.5			& 86 \%	& 		& 		& 7 \%	& 7 \%		& 			\\
\end{tabular}
\end{ruledtabular}
\caption{\label{charm_jpp_table}Calculated masses, $c\overline{c}$ and meson-meson probabilities for $J^{++}$ charmonium-like states. A missing entry
means that the corresponding component gives negligible (i.e.\ inferior to $1\%$) or no contribution to the state.}
\end{table*}

It is illustrative to compare these results with the $c\overline{c}$ masses in
Table~\ref{cortable} obtained with the Cornell potential. A glance at these tables
makes clear that the presence of the thresholds gives rise to attraction in
the sense that the resulting masses are reduced with respect the corresponding
Cornell $c\overline{c}$ masses. For the lowest-lying $0^+(J^{++})$ states ($
J=0,1,2$) there is a very small mass difference indicating an almost
negligible attraction for these states. This is understood for the thresholds
are far above in energy ($\geq200$~MeV) so that no significant mixing occurs
(less than $1\%$ meson-meson probability).

The situation is completely altered for the first excited $0^+(J^{++})$ states. Thus, the fitting of
the first excited $1^{++}$ resonance, $\chi_{c_{1}}(3872)$
with a measured mass of $3871.69\pm0.17$~MeV, requiring\ a mass
reduction of $81$~MeV with respect to the Cornell $c\overline{c}$ mass,
implies a very strong mixing, $99\%$ of $D\overline{D}^*$ component,
whereas for the $0^{++}$ and $2^{++}$ states the predicted mixing is about
$40\%$ (mainly from $D_{s}\overline{D}_{s})$ and $15\%$ (shared by
$D_{s}\overline{D}_{s}$ and $D^*\overline{D}^*)$ respectively, with
corresponding mass reductions of $33$~MeV and $20$~MeV.

It is amazing that these $(0,2)^{++}$ mass predictions are in
complete agreement with data regarding their positions with respect to
the $D_{s}\overline{D}_{s}\ $threshold, both below it. Moreover, their calculated
numerical values are pretty close to the measured ones. So, the calculated
$2^{++}$ mass, $3933.5$~MeV, is very close to that of the experimental
resonance $\chi_{c_{2}}(3930)$: $3927.2\pm2.6$~MeV. And the $0^{++}$
calculated mass, $3920.4$~MeV, is consistent with the ones of the experimental
candidates: $\chi_{c_{0}}(3860)$, with a measured mass of $3862_{-32-13}%
^{+26+40}$~MeV, and $X(3915)$ with a measured mass of $3918.4\pm1.9$~MeV,
although in this last case the assignment to a $2^{++}$ state cannot be
completely ruled out, see \cite{PDG20} and references therein. This suggests
that further mass corrections for these sates as the ones due to
spin-dependent terms in the $c\overline{c}$ potential,
or to one pion exchange in the meson-meson potential,
or those taking into account the effect of the lower threshold $D\overline
{D}$, or the deviations from the assumption of the same
values of the mixing potential parameters for all the thresholds,
are either small and might be implemented perturbatively, or have been
partially taken into account through the effectiveness of the
parameters of the mixing potential.

It should also be emphasized that our nonperturbative formalism provides us
with the meson wave functions in terms of their $c\overline{c}$ and
meson-meson components.

\begin{figure}
\includegraphics{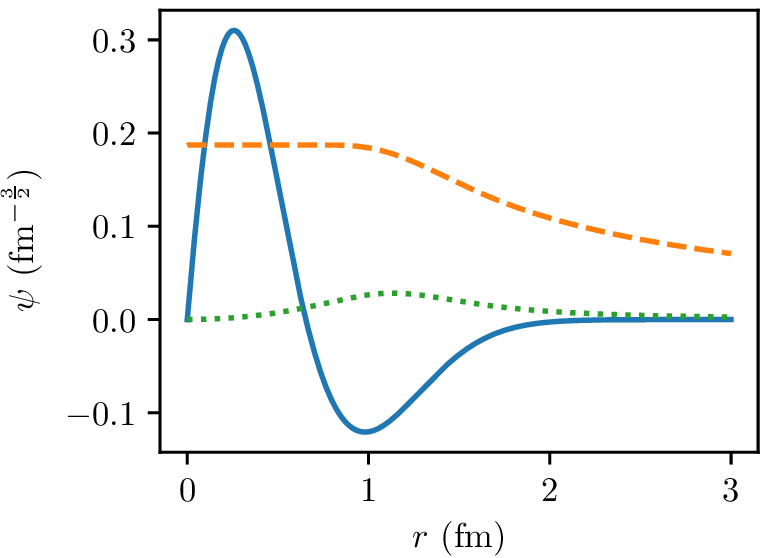}
\caption{Radial wave function of the calculated $0^+(1^{++})$ state with a mass of $3871.7$~MeV.
$c\overline{c}(2\,{^{3}\!p}_{1})$, $D \overline{D}^*(l_{D \overline{D}^*}=0)$
and $D \overline{D}^*(l_{D \overline{D}^*}=2)$ components are drawn with a solid, dashed and dotted line respectively.}
\label{xc13872}
\end{figure}

For $\chi_{c_{1}}(3872)$ the radial $c\overline{c}$
and $D\overline{D}^*$ ($l_{D\overline{D}^*}=0,2$) wave function
components are plotted in Fig.~\ref{xc13872}.

A look at this figure makes clear the prevalence of the $D\overline{D}^*$
channel with $l_{D\overline{D}^*}=0$ for
distances beyond $2$~fm. As the estimated Cornell rms radius for $D$ is about
$0.54$~fm we may conclude that $\chi_{c_{1}}(3872)$, with a
calculated rms radius of $26.17$~fm is at large distances a loose hadromolecular
state. At short distances, though, the $(2\,{^{3}\!p}_{1})$
$c\overline{c}$ component, with a rms radius of $1.01$~fm plays a role at
least as prominent as the $D\overline{D}^*$ one, see Fig.~\ref{xc13872}. These
features are quite in line with the indications from phenomenology requiring a
$c\overline{c}$ component to give proper account of short distance properties.

\begin{figure}
\includegraphics{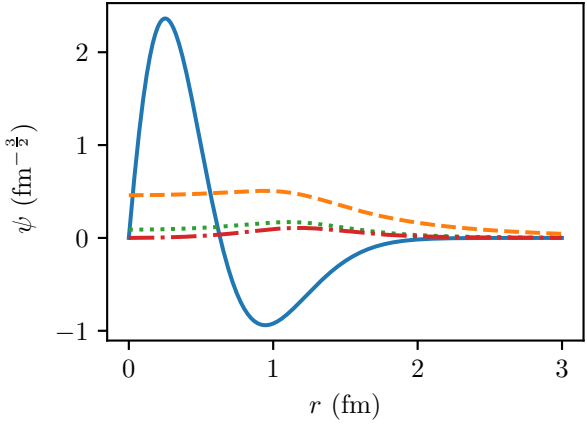}
\caption{\label{xc03915}Radial wave function of the calculated $0^+(0^{++})$ state with a mass
of $3920.4$~MeV. $c\overline{c}(2\,{^{3}\!p}_{0})$, $D_s \overline{D}_s(l_{D_s \overline{D}_s}=0)$,
$D^* \overline{D}^*(l_{D^* \overline{D}^*}=0)$ and $D^* \overline{D}^*(l_{D^* \overline{D}^*}=2)$
components are drawn with a solid, dashed, dotted and dash-dotted line respectively.}
\end{figure}

For the calculated $0^{++}$ state the radial wave function is drawn in
Fig.~\ref{xc03915}.

As can be checked, the wave function with a rms radius of $1.26$
fm is made mainly of $c\overline{c}$ and
$D_{s}\overline{D}_{s}$ with a $59\%$ and $37\%$ probability respectively.
This indicates a dominant
$D\overline{D}$ strong decay mode from $c\overline{c}$ as it is experimentally
the case for $\chi_{c_{0}}(3860)$. On the other hand, a $J/\psi\,\omega$ decay
mode may get a significant contribution from $D_{s}\overline{D}_{s}$ since it
is OZI allowed through the small $s\overline{s}$ content of $\omega$. This
could cause this mode to be also a dominant one as it is experimentally the
case for $X(3915)$. Hence, it could be that $\chi_{c_{0}}(3860)$ and $X(3915)$
are just the same resonance observed through two different decay modes.

\begin{figure}
\includegraphics{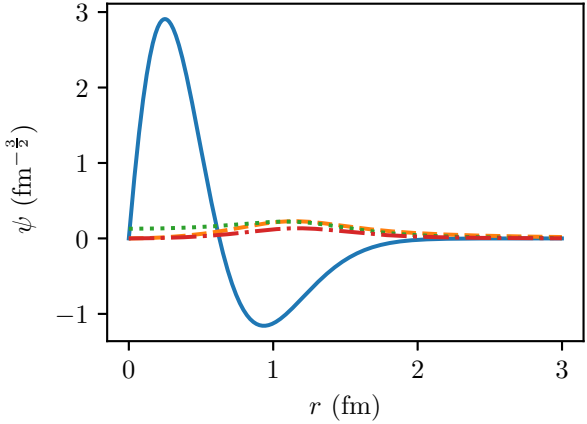}
\caption{\label{xc23930}Radial wave function of the calculated $0^+(2^{++})$ state with a
mass of $3933.5$~MeV.  $c\overline{c}(2\,{^{3}\!p}_{2})$, $D_s \overline{D}_s(l_{D_s \overline{D}_s}=2)$,
$D^* \overline{D}^*(l_{D^* \overline{D}^*}=0)$ and $D^* \overline{D}^*(l_{D^* \overline{D}^*}=2)$
components are drawn with a solid, dashed, dotted and dash-dotted line respectively.}
\end{figure}

As for the calculated $2^{++}$ state, the wave function, with a rms radius of
$1.06$~fm, is plotted in Fig.~\ref{xc23930}. It is mostly that of the $c\overline{c}$
component. This is in accord with a very dominant $D\overline{D}$ strong decay
mode as it is experimentally the case for $\chi_{c_{2}}(3930)$.

Certainly these qualitative arguments on the dominant strong decay modes
should be supported by trustable and predictive quantitative calculations.
As mentioned before, the
development of a consistent formalism for the calculation of the decay widths
to open flavor meson-meson states, which is out of the scope of this article,
is in progress. One should keep in mind though that the dearth of current
detailed quantitative decay data for comparison will be a serious drawback to
test it. We strongly encourage experimental efforts along this line.

Regarding electromagnetic radiative transitions, although important
progress for the accurate calculation of decays from the $c\overline{c}$
component has been reported \cite{Bru20}, a reliable and consistent
calculation incorporating the meson-meson contribution as well is lacking. We
encourage a theoretical effort along this line.

One can do better, as we show next, for leptonic decays from the low-lying $1^{--}$ states since the decay widths
depend on the wave function at the origin and the contribution from meson-meson components is suppressed as
they are not in $s$-wave, see Table~\ref{quanum}.

The calculated $1^{--}$ spectrum of states is listed in Table~\ref{charm_omm_table}.

\begin{table*}
\begin{ruledtabular}
\begin{tabular}{ccdddddd}
$J^{PC}$	& Mass (MeV)	& c\overline{c}	& D \overline{D}	& D \overline{D}^*
	& D_s \overline{D}_s	& D^* \overline{D}^*	& D_s \overline{D}_s^*	\\
\hline
\multirow{4}*{$1^{--}$}	& 3082.4			& 100 \%	& 		& 		& 		& 			& 			\\
		& 3664.2			& 95 \%	& 4 \%	& 1 \%	& 		& 			& 			\\
		& 3790.2			& 97 \%	& 		& 2 \%	& 1 \%	& 			& 			\\
		& 4071.0			& 64 \%	&		& 		& 		& 			& 36  \%		\\
\end{tabular}
\end{ruledtabular}
\caption{\label{charm_omm_table}Calculated masses, $c\overline{c}$ and meson-meson probabilities for
$1^{--}$ charmonium-like states. A missing entry means that the corresponding component gives negligible
(i.e.\ inferior to $1\%$) or no contribution to the state.}
\end{table*}

Again, a comparison with the $c\overline{c}$ masses in Table~\ref{cortable} makes clear
that the presence of the thresholds gives rise to attraction. As it was the
case for $J^{++}$, the lowest state, lying far below the lowest threshold, has
no mixing at all being the $1s$ $c\overline{c}$ state. A pretty small mixing
is present for the next two higher states that can be mostly assigned to the
$2s$ $(95\%)$ and $1d$ $(97\%)$ $c\overline{c}$
states respectively. It is worth to mention that for the $1d$ state with a
Cornell $c\overline{c}$ mass of $3795.8$~MeV, the $D\overline{D}$ threshold
lying $66$~MeV below does not produce enough attraction to bring the state below threshold.

\begin{figure}
\includegraphics{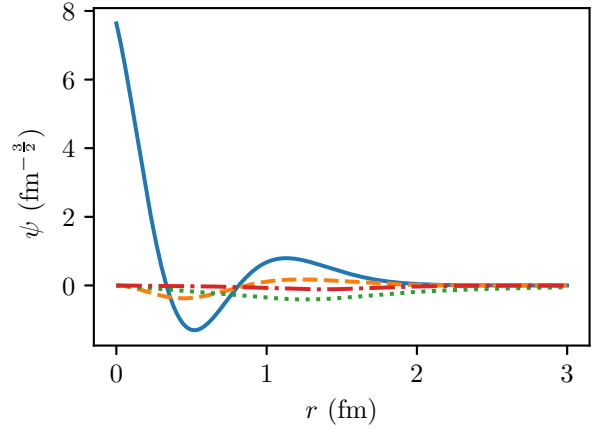}
\caption{\label{psi4040}Radial wave function of the calculated $0^{-}(1^{--})$ state with
a mass of $4071$~MeV.  $c\overline{c}(3\,{^{3}\!s}_{1})$, $c\overline{c}(2\,{^{3}\!d}_{1})$
and $D_s \overline{D}_s^*(l_{D_s \overline{D}_s^*}=1)$ components are drawn with a solid,
dashed and dotted line respectively.}
\end{figure}

The first state with a significant mixing, $36\%$ of $D_{s}\overline{D}_{s}^*$,
is predicted at $4071$~MeV and contains a $60\%$ of $3s$
$c\overline{c}$ and a $4\%$ of $2d$ $c\overline{c}$ as well. Its wave function
is drawn in Fig.~\ref{psi4040}.

In this case the vicinity of the $D_{s}\overline{D}_{s}^*$ threshold at
$4080$~MeV to the $3s$ $c\overline{c}$ Cornell mass at $4097$~MeV produces
sufficient attraction to bring the state below threshold, in agreement with
data under its assignment to the $\psi(4040)$ resonance with a
measured mass of $4039\pm1$~MeV. Furthermore the expected dominant decay
modes, ($D\overline{D},D\overline{D}^*,D_{s}\overline{D}_{s},D^*\overline{D}^*$) from $c\overline{c}$, and
$D_{s}\overline{D}_{s}\,\gamma$ from $D_{s}\overline{D}_{s}^*$, are in
perfect accord with the ones observed from $e^{+}e^{-}\rightarrow \text{hadrons}$.

As for the well-measured leptonic width
\begin{equation}
\left(\Gamma\left(\psi(4040)\rightarrow e^{+}e^{-}\right)\right)
_\text{Expt}=0.86\pm0.07\text{~KeV},
\end{equation}
we can trustfully predict the ratios
\begin{subequations}\label{thratios}
\begin{equation}
\frac{\Gamma_{\psi(4040)\rightarrow e^{+}e^{-}}^\text{Theor}}{\Gamma_{\psi(1s)\rightarrow e^{+}e^{-}}^\text{Theor}}
=\frac{\abs{R_{\psi(4040)}(0)}^{2}}{\abs{R_{\psi(1s)}(0)}^{2}}\frac{M_{\psi(1s)}^{2}}{M_{\psi(4040)}^{2}}\\
\approx 0.18
\end{equation}
and
\begin{equation}
\frac{\Gamma_{\psi(4040)\rightarrow e^{+}e^{-}}^\text{Theor}}{\Gamma_{\psi(2s)\rightarrow e^{+}e^{-}}^\text{Theor}}
=\frac{\abs{R_{\psi(4040)}(0)}^{2}}{\abs{R_{\psi(2s)}(0)}^{2}}\frac{M_{\psi(2s)}^{2}}{M_{\psi(4040)}^{2}}\\
\approx 0.43
\end{equation}
\end{subequations}
to be compared to
\begin{subequations}\label{expratios}
\begin{equation}
\frac{\Gamma_{\psi(4040)\rightarrow e^{+}e^{-}}^\text{Expt}}{\Gamma_{\psi(1s)\rightarrow e^{+}e^{-}}^\text{Expt}}
=0.15\pm0.03
\end{equation}
and
\begin{equation}
\frac{\Gamma_{\psi(4040)\rightarrow e^{+}e^{-}}^\text{Expt}}{\Gamma_{\psi(2s)\rightarrow e^{+}e^{-}}^\text{Expt}}
=0.37\pm0.07 .
\end{equation}
\end{subequations}

Hence, our results agree with data within the experimental intervals. The
reason for this agreement has to do with the reduced probability of the $3s$
$c\overline{c}$ component, $60\%$, induced by the mixing with the
$D_{s}\overline{D}_{s}^*$ threshold. This mixing is also responsible for
the $4\%$ of $2d$ $c\overline{c}$ component. This small (big) $2d$ ($3s$)
probability could be increased (decreased) if a tensor interaction were
incorporated as a correction term to the Cornell potential. Maybe the bias we
observe in our results, both agreeing with the maximum allowed experimental
values, is an indication in this sense. In any case a modest additional
probability reduction of the $3s$ $c\overline{c}$ component should be expected.

It is worth to mention that the explanation of the leptonic width for
$\psi(4040)$ has been linked in the literature to that of
$\psi(4160)$ through a very significant $s$-$d$ mixing
\cite{Bad09}. Our results do not support this idea. Instead the $D_{s}\overline{D}_{s}^*$ -- $c\overline{c}(3s)$
mixing appears to be the main physical mechanism underlying the $\psi(4040)$ decay to
$e^{+}e^{-}$.

Unfortunately, at the current stage of our diabatic development
we cannot properly evaluate $\psi(4160)$, the main reason being
that the dominant Cornell $2d$ $c\overline{c}$ state lies only $100$~MeV below
the first $s$-wave $1^{--}$ threshold, $D\overline{D}_{1}$, which is composed
of two overlapping thresholds, $D\overline{D}_{1}(2420)$ and
$D\overline{D}_{1}(2430)$, the last one with a large width.
Quite presumably this double threshold gives a significant contribution by
itself to the leptonic width of $\psi(4160)$.

This current limitation applies as well to the description of unconventional
states with masses above $4.1$~GeV such as $\psi(4260)$ lying
close below the $D\overline{D}_{1}$ double threshold, or $\psi(4360)$ and $\psi(4415)$ lying close below a multiple
threshold at $4429$~MeV. The same limitation applies for $J^{++}$ states. Work along this line is in progress.

\section{\label{sec6}Summary and conclusions}

A general formalism for a unified description of conventional and
unconventional heavy-quark meson states has been\ developed and successfully
applied to isoscalar $J^{++}$ and $1^{--}$ charmonium-like states with masses
below $4.1$~GeV.

The formalism adapts the diabatic approach, widely used in molecular physics
to tackle the configuration mixing problem, to the study of heavy-quark meson
states involving quark-antiquark as well as meson-meson components. A great
advantage of using this approach, against the Born-Oppenheimer ({\BO})
approximation commonly used for heavy-quark mesons, is that the bound states
are expanded in terms of $Q\overline{Q}$ and meson-meson configurations
instead of the mixed configurations that correspond to the ground and
excited states of the light fields. Then instead of being forced to use a
single channel approximation to solve the bound state problem as in {\BO},
what in practice is equivalent to neglect the configuration mixing, one
can write a treatable multichannel {\schr} equation where the
interaction between configurations is incorporated through a diabatic
potential matrix. Moreover, the diagonal and off-diagonal elements of this
potential matrix can be directly related to the static energies obtained from
\textit{ab initio} quenched (only $Q\overline{Q}$ or meson-meson
configuration) and unquenched ($Q\overline{Q}$ and meson-meson configurations)
lattice calculations. This connection defines the diabatic
approach in QCD.

It is worth to emphasize that this approach goes also beyond the incorporation
of hadron loop corrections to the {\BO} scheme that have been used sometimes in
the literature to deal with unconventional charmonium-like mesons. Indeed, the
diabatic bound state wave functions, given in terms of quark-antiquark and
meson-meson components, allow for a complete nonperturbative evaluation of
observable properties.

This theoretical framework has been tested in the charmonium-like meson sector
where there is compelling evidence of the existence of mixed-configuration
states, in particular the very well-established $0^{+}(1^{++})$
resonance $\chi_{c_{1}}(3872)$ that we use to fix our
parametrization of the mixing potential.

Although a complete (at all energies) spectral description would require
additional theoretical refinements, as for example the incorporation of
threshold widths, the results obtained for states with mass below $4.1$~GeV,
for which the significant thresholds are very narrow, are encouraging. All the
mass values are well reproduced and their locations with respect to the
thresholds correctly predicted making clear the $c\overline{c}$ -- threshold
attraction. This points out to the diabatic
approach as an appropriate framework for a unified and complete
nonperturbative description of heavy-quark meson states.

\begin{acknowledgments}
This work has been supported by MINECO of Spain and EU Feder Grant No.~FPA2016-77177-C2-1-P, by SEV-2014-0398,
by EU Horizon 2020 Grant No.~824093 (STRONG-2020) and by PID2019-105439GB-C21. R.~B. acknowledges a FPI
fellowship from MICIU of Spain under Grant No.~BES-2017-079860.
\end{acknowledgments}

\appendix

\section{\label{apdxADT}Adiabatic-to-diabatic transformation}

As the light field eigenstates $\ket{\zeta_{i}(\vb*{r})}$ form a complete orthonormal set whatever
the value of $\vb*{r}$, we can express (we use hereby Einstein notation so that a sum over repeated indices is understood)
\begin{equation}
\ket{\zeta_{j}(\vb*{r}_0)} = \ket{\zeta_{i}(\vb*{r})} A_{i j}(\vb*{r},\vb*{r}_0)
\label{cb}
\end{equation}
where $A_{i j}(\vb*{r},\vb*{r}_0)$ is a change of basis unitary matrix defined formally by
\begin{equation}
A_{i j}(\vb*{r},\vb*{r}_0) \equiv \braket{\zeta_{i}(\vb*{r})}{\zeta_{j}(\vb*{r}_0)} .
\label{ADTM1}
\end{equation}
This matrix, which is a function of the coordinate $\vb*{r}$ and depends parametrically on the fixed point $\vb*{r}_0$, is referred to in this context
as the \emph{Adiabatic-to-Diabatic Transformation matrix} (ADT matrix). Let us examine the conditions to be satisfied by the ADT matrix
for the adiabatic and diabatic expansions to be equivalent \cite{Bae06}.

Let us begin by inserting \eqref{cb} in the diabatic expansion \eqref{DSTATE} and comparing with \eqref{BOSTATE}.
We thus see that  the ADT matrix transforms the diabatic wave function in the adiabatic one:
\begin{equation}
\psi_{i}(\vb*{r}) = A_{i j}(\vb*{r}, \vb*{r}_0) \widetilde{\psi}_{j}(\vb*{r},\vb*{r}_0).
\label{adtdef}
\end{equation}
If we now plug \eqref{adtdef} into Eq.~\eqref{BSE} and multiply on the left by $\mathrm{A}^\dagger$ we obtain
\begin{widetext}
\begin{equation}
\bqty{-\frac{\hbar^2}{2\mu_{Q\overline{Q}}}A^\dagger_{i k}((\grad + \vb*{\tau})^2)_{k l} A_{l j} +
(A^\dagger_{i k}V_k A_{k j} - \delta_{i j} E)} \widetilde{\psi}_{j} = 0,
\label{transtep}
\end{equation}
where we have momentarily dropped the arguments $\vb*{r}$
and $\vb*{r}_0$ to simplify the notation. Using
\begin{equation}
\begin{split}
((\grad + \vb*{\tau})^2)_{k l} A_{l j} \widetilde \psi_{j} =&
(\delta_{k m} \grad + \vb*{\tau}_{k m}) \vdot  (\delta_{m l} \grad + \vb*{\tau}_{m l}) A_{l j} \widetilde \psi_{j} \\
=& (\delta_{k m} \grad + \vb*{\tau}_{k m}) \vdot (A_{m j} \grad  + (\grad A)_{m j} + \vb*{\tau}_{m l} A_{l j} )\widetilde \psi_j \\
=& [A_{k j} \laplacian + 2 (\grad A)_{k j} \vdot \grad + (\laplacian A)_{k j}
+(\div \vb*{\tau})_{k l} A_{l j} + 2 \vb*{\tau}_{k l} \vdot (\grad A)_{l j}+ 2 \vb*{\tau}_{k l}
\vdot A_{l j} \grad + (\vb*{\tau}^2)_{k l} A_{l j}] \widetilde \psi_j \\
=& [A_{k j} \laplacian + ((\grad A)_{k j} + \vb*{\tau}_{k l} A_{l j})\vdot \grad
+ (\delta_{k l} \grad + \vb*{\tau}_{k l})\vdot((\grad A)_{l j} + \vb*{\tau}_{l m} A_{m j})] \widetilde\psi_j ,
\end{split}
\end{equation}
we can expand the kinetic term as
\begin{equation}
A^\dagger_{i k}((\grad + \vb*{\tau})^2)_{k l} A_{l j} = \delta_{i j} \laplacian + A^\dagger_{i k} ((\grad A)_{k j} + \vb*{\tau}_{k l} A_{l j})\vdot \grad
+ (A^\dagger_{i l}  \grad + A^\dagger_{i k}  \vb*{\tau}_{k l})\vdot((\grad A)_{l j} + \vb*{\tau}_{l m} A_{m j}).
\label{kinexpand}
\end{equation}
Therefore, as in the diabatic representation the kinetic term is diagonal, the ADT matrix must satisfy the first order differential equation
\begin{equation}
\grad A_{i j}(\vb*{r}, \vb*{r}_0) + \vb*{\tau}_{i k}(\vb*{r}) A_{k j}(\vb*{r}, \vb*{r}_0) = 0 ,
\label{adtmatdiff}
\end{equation}
where we have restored the arguments $\vb*{r}$ and $\vb*{r}_0$.

Eq. \eqref{adtmatdiff}, together with the boundary condition $A_{ij}(\vb*{r}_0, \vb*{r}_0)=\delta_{i j}$, determines uniquely the ADT
matrix for every point in configuration space, if the NACTs are well-behaved. If otherwise the NACTs present singularities,
the ADT matrix may be multi-valued \cite{Bae06}. We will not examine this latter possibility here.

Substituting \eqref{kinexpand}-\eqref{adtmatdiff}, Eq.~\eqref{transtep} becomes
\begin{equation}
\bqty{-\frac{\hbar^2}{2\mu_{Q\overline{Q}}}\delta_{i j} \laplacian +
(A^\dagger_{i k}(\vb*{r}, \vb*{r}_0) V_k(\vb*{r}) A_{k j}(\vb*{r}, \vb*{r}_0) - \delta_{i j} E)} \widetilde \psi_{j}(\vb*{r}, \vb*{r}_0) = 0,
\end{equation}
\end{widetext}
which can be recognized as the diabatic {\schr} equation \eqref{DEQ} by requiring
\begin{equation}
A^\dagger_{i k}(\vb*{r}, \vb*{r}_0) V_k(\vb*{r}) A_{k j} (\vb*{r}, \vb*{r}_0) = V_{i j}(\vb*{r}, \vb*{r}_0).
\label{adtmatdiag}
\end{equation}
This requirement tells us that the ADT matrix diagonalizes the diabatic potential matrix,
and that the eigenvalues of the diabatic potential matrix are
then the unquenched static energies $V_i(\vb*{r})$.

It is thus proved that the diabatic and adiabatic expansions are completely equivalent, so that the NACTs together
with the unquenched static energies carry the same amount of physical information as the diabatic potential matrix.

\section{\label{apdxcpar}$C$-parity of meson-meson states}

Although heavy-light mesons do not have definite $C$\nobreakdash-parity
nor $G$\nobreakdash-parity, meson-meson configurations with $I=0$ can be rearranged
in combinations with definite $C$\nobreakdash-parity. To build these combinations let us
start by observing the action of $C$\nobreakdash-parity on some heavy-light
meson state $M$ made of a light quark $q$ and a heavy antiquark $\overline{Q}$:
\begin{equation}
C \ket{M} = (-1)^{l_M + s_M} \ket{\overline{M}} 
\end{equation}
where $l_M$ and $s_M$, the internal orbital angular momentum and internal spin of the meson, are given
in terms of the $q\overline{Q}$ relative orbital angular momentum and total spin respectively. Next we
consider the action of $C$\nobreakdash-parity on the isospin singlet state formed by a $q\overline{Q}$ meson $M_1$
and a $Q\overline{q}$ meson $\overline{M}_2$
\begin{equation}
C \ket*{M_1 \overline{M}_2}_0 = (-1)^{l_{M_1} + s_{M_2} + l_{\overline{M}_2} + s_{\overline{M}_2}} \ket*{\overline{M}_1 M_2}_0.
\label{sign1}
\end{equation}
We now exchange the positions and spin labels of the mesons in $\ket*{\overline{M}_1 M_2}_0$, thus obtaining an additional sign:
\begin{equation}
\ket*{\overline{M}_1 M_2}_0 = (-1)^{l_{M_1\overline{M}_2} + s_{M_1\overline{M}_2} + j_{M_1} + j_{\overline{M}_2}} \ket*{M_2 \overline{M}_1}_0
\label{sign2}
\end{equation}
where $j_M$ is the total spin of the meson given by the sum of the meson internal orbital angular momentum and spin.
Note that the factor $(-1)^{l_{M_1\overline{M}_2}}$ comes from the exchange of the positions and the factor
$(-1)^{s_{M_1\overline{M}_2} + j_{M_1} + j_{\overline{M}_2}}$ comes from the exchange of the spin labels.
Then substituting \eqref{sign2} in \eqref{sign1} we obtain
\begin{equation}
C \ket*{M_1 \overline{M}_2}_0 = \mathcal{C}_{M_1 \overline{M_2}} \ket*{M_2 \overline{M}_1}_0
\label{eqC}
\end{equation}
where
\begin{equation}
\mathcal{C}_{M_1 \overline{M}_2}\equiv (-1)^{l_{M_1 \overline{M}_2} + s_{M_1 \overline{M}_2}
+ l_{M_1} + l_{\overline{M}_2} + s_{M_1} + s_{\overline{M}_2} + j_{M_1} + j_{\overline{M}_2}}.
\end{equation}
From Eq.~\eqref{eqC} it is then straightforward to prove that the states
\begin{equation}
\ket*{M_1 \overline{M}_2}_\pm \equiv \frac{1}{\sqrt{2}}\pqty{\ket*{M_1 \overline{M_2}}_0 \pm
\mathcal{C}_{M_1 \overline{M}_2} \ket*{M_2 \overline{M}_1}_0}
\end{equation}
have definite $C$\nobreakdash-parity:
\begin{equation}
\begin{split}
C \ket*{M_1\overline{M}_2}_{\pm} &= \frac{1}{\sqrt{2}} \pqty{C \ket*{M_1\overline{M}_2}_0
\pm \mathcal{C}_{M_1 \overline{M_2}} C \ket*{M_2\overline{M}_1}_0} \\
&= \frac{1}{\sqrt{2}} \pqty{\mathcal{C}_{M_1 \overline{M_2}} \ket*{M_2\overline{M}_1}_0 \pm \ket*{M_1\overline{M}_2}_0} \\
&= \pm \frac{1}{\sqrt{2}} \pqty{\ket*{M_1\overline{M}_2}_0 \pm \mathcal{C}_{M_1 \overline{M_2}} \ket*{M_2\overline{M}_1}_0} \\
&\equiv \pm \ket*{M_1\overline{M}_2}_{\pm} ,
\end{split}
\end{equation}
where we have used the fact that $\mathcal{C}_{M_1 \overline{M_2}}^2 = (\pm)^2 = 1$.

\section{\label{apdxvar}Variational method}

To solve the {\schr} equation we use a variational method, its essence being that given a Hamiltonian $H$ defined over 
a Hilbert space $\mathbb{H}$, and defining the functional
\begin{equation}
\mathcal{F}[\varphi] \equiv \frac{\mel{\varphi}{H}{\varphi}}{\braket{\varphi}},
\end{equation}
where $\ket{\varphi} \in \mathbb{H} \setminus \Bqty{0}$ is some non-null vector in the Hilbert space, the eigenvectors
of $H$ correspond to stationary points of $\mathcal{F}$, and the values of the functional on those stationary points are the corresponding
eigenvalues:
\begin{equation}
H \ket{\psi_n} = E_n \ket{\psi_n} \iff \var{\mathcal{F}[\psi_n]} = 0 \; \land \; \mathcal{F}[\psi_n] = E_n .
\end{equation}
To show this, we first reduce the variational problem of finding the stationary points of $\mathcal{F}$ to an algebraic problem by expanding the
state $\ket{\varphi}$ in terms of an orthonormal basis of $\mathbb{H}$
\begin{equation}
\ket{\varphi} = \sum_i \varphi_i \ket{e_i},
\end{equation}
so that the functional $\mathcal{F}$ becomes an ordinary function of the coordinates
\begin{equation}
\mathcal{F}[\varphi] \rightarrow \mathcal{F}(\varphi_1,\varphi_2,\dots) = \frac{\sum_{j, k} \varphi^*_j H_{j k} \varphi_k}{\sum_j \abs{\varphi_j}^2},
\label{funcoord}
\end{equation}
where we have introduced the Hamiltonian matrix elements
\begin{equation}
H_{i j} \equiv \mel{e_i}{H}{e_j}.
\end{equation}
Second, we determine which values of the coordinates $\varphi_i$ correspond to stationary points of $\mathcal{F}$. With the
functional derivative becoming an ordinary one, the stationary points are found as the solutions of
\begin{equation}
\fdv{\mathcal{F}}{\varphi} \rightarrow \pdv{\mathcal{F}}{\varphi_i} = 0 .
\end{equation}
for every $i$. Using \eqref{funcoord} and expanding the derivatives we obtain
\begin{equation}
\frac{2}{\sum_j \abs{\varphi_j}^2} \pqty{\sum_j \varphi^*_j H_{j i} - \varphi^*_i
\pqty{\frac{\sum_{j,k}\varphi^*_j H_{j k} \varphi_k}{\sum_j \abs{\varphi_j}^2}}} = 0
\end{equation}
or equivalently
\begin{equation}
\sum_j H_{i j} \varphi_j = \mathcal{F}(\varphi_1,\varphi_2,\dots) \varphi_i .
\label{varsol}
\end{equation}
Eq.~\eqref{varsol} is nothing but the characteristic equation for $H$ in the matrix representation provided by $\Bqty{\ket{e_i}}$. Therefore it is
proved that the states $\ket{\varphi}$ corresponding to stationary points of $\mathcal{F}$ are also eigenstates of $H$. Moreover,
Eq.~\eqref{varsol} shows that the value of the functional $\mathcal{F}$ at the stationary point is precisely
the corresponding energy eigenvalue.

Technically speaking, the results presented here are analytically valid only when using a complete, i.e.\ infinite, orthonormal set. Since in realistic
applications one employs a limited set, the correspondence drawn here is only approximate and so are the energies and eigenstates
obtained with the variational method.

A shortcoming of the variational method is that the degree of approximation is not known \textit{a priori}. To assure this not to be any problem
we choose an appropriate orthonormal set of states reflecting some of the properties of the physical states and employ a very high number of states
in the set.

\section{\label{apdxlag}Laguerre associated polynomials}

For the solution of the {\schr} equation with a spherical potential a natural (physical) choice for a basis describing the radial wave function is the one
of associated Laguerre polynomials. These are explicitly defined by
\begin{equation}
L_n^k (x) = \sum_{i=0}^n \frac{n!}{i!} \binom{n + k}{n - i} (-x)^i
\end{equation}
where $\binom{k + n}{n - i}$ is a binomial coefficient, and form an orthogonal basis set of $\mathbb{L}^2(0,\infty)$ with
weighting function $x^k e^{-x}$:
\begin{equation}
\int_0^\infty \dd{x} x^k e^{-x} L_n^k (x) L_m^k (x) =  \frac{(n+k)!}{n!} \delta_{n m} .
\label{lagorto}
\end{equation}
More precisely, the solutions of the spherical {\schr} equation factorize in a spherical harmonic and
a radial wave function as
\begin{equation}
\psi_{E, l}^m(\vb*{r}) = u_{E,l}(r)  Y_l^m(\vu*{r}),
\end{equation}
where the radial wave function $u_{E,l}(r)$ has well-known asymptotic behaviors. For bound states, these are
\begin{equation}
u_{E,l}(r) \overset{r\to0}{\sim} \pqty{\frac{r}{\lambda_E}}^l \qand u_{E,l}(r) \overset{r\to\infty}{\sim} e^{-\frac{r}{2 \lambda_E}}
\end{equation}
where $\lambda_E$ is some length scale that may depend on the bound state mass $E$. Knowing this we can write in general the radial wave function as
\begin{equation}
u_{E,l}(r) = \lambda_E^{-\frac{3}{2}} \pqty{\frac{r}{\lambda_E}}^l \mathcal{U}_{E,l} \pqty{\frac{r}{\lambda_E}} e^{-\frac{r}{2 \lambda_E}}
\label{gen}
\end{equation}
where $\mathcal{U}_{E,l}\pqty{\frac{r}{\lambda_E}}$ must be some scalar function that does
not vanish for $r \to 0$ and diverges at most as a power of $r$ for
$r \to \infty$. Then, normalization of the radial wave function
\begin{equation}
\int_0^{\infty} \dd{r} r^2 u_{E,l} (r) u_{E',l} (r) = \delta_{E E'},
\end{equation}
reads
\begin{equation}
\begin{split}
\int_0^{\infty} \frac{\dd{r}}{\sqrt{\lambda_E \lambda_{E'}}}& \frac{r^{2 l + 2}}{(\lambda_E \lambda_{E'})^{l+1}}
e^{-r \frac{\lambda_E + \lambda_{E'}}{ 2 \lambda_E  \lambda_{E'}}} \\
& \times \mathcal{U}_{E,l} \pqty{\frac{r}{\lambda_E}} \mathcal{U}_{E',l} \pqty{\frac{r}{\lambda_{E'}}} = \delta_{E E'}.
\end{split}
\label{genorto}
\end{equation}
We can now compare this result with the one resulting from \eqref{lagorto} when substituting $x\to r/\lambda$, with $\lambda$ being some
constant with dimensions of length. We obtain
\begin{equation}
\int_0^\infty \frac{\dd{r}}{\lambda} \pqty{\frac{r}{\lambda}}^k e^{-\frac{r}{\lambda}} L_n^k \pqty{\frac{r}{\lambda}}
L_m^k \pqty{\frac{r}{\lambda}} =  \frac{(n+k)!}{n!} \delta_{n m} ,
\end{equation}
that corresponds to \eqref{genorto} with $2l+2=k$ and $\lambda_E=\lambda_{E'}=\lambda$ up to a normalization factor.

It is then quite clear that the most natural choice for a basis is
\begin{equation}
e_{n,l}^{m}(\vb*{r}) =  N_{n,l} \pqty{\frac{r}{\lambda}}^l L_n^{2l+2}\pqty{\frac{r}{\lambda}} e^{-\frac{r}{2 \lambda}} Y_l^m(\vu*{r})
\label{thebase}
\end{equation}
being $N_{n,l}$ the normalization factor
\begin{equation}
N_{n,l} \equiv \bqty{\lambda^3 \frac{(n + 2l + 2)!}{n!}}^{-\frac{1}{2}}
\label{thenorm}
\end{equation}
such that the basis is orthonormal:
\begin{equation}
\braket*{e_{n,l}^{m}}{e_{n',l'}^{m'}}=\delta_{n n'} \delta_{l l'} \delta_{m m'}.
\end{equation}

The basis defined by \eqref{thebase}-\eqref{thenorm} is expected to provide a reasonable description of the physical eigenstates
as long as the scale $\lambda$ is roughly of the same order that the physical scales $\lambda_E$ involved and the number
$n_\text{max}$ of polynomials used in the calculation is high enough.

Given that any numerical calculation of this kind is performed on a discretized ($r_{n} - r_{n-1} = \delta$)
and limited ($r_{n}\le r_\text{max}$) radial configuration space, the hyperparameters involved in this scheme are:
\begin{itemize}
\item $\delta$: the discretization step of $r$;
\item $r_\text{max}$: the maximum integration radius;
\item $\lambda$: the length scale in the associated Laguerre basis;
\item $n_\text{max}$: the number of associated Laguerre polynomials used.
\end{itemize}
In this work we use $\delta=10^{-3}$~fm, $r_\text{max}=150$~fm, $\lambda = 0.2$~fm and $n_\text{max}=150$.

Note that when doing numerical calculations following this procedure one should always check stability of the results under changes
of these hyperparameters, keeping in mind that convergence with higher values of $\lambda$ and $n_\text{max}$ demands bigger values for
$r_\text{max}$, and that $\delta$ should always be small enough in order to keep numerical integration errors under control.

\bibliography{diabatic-bib}

\begin{thebibliography}{26}%
\makeatletter
\providecommand \@ifxundefined [1]{%
 \@ifx{#1\undefined}
}%
\providecommand \@ifnum [1]{%
 \ifnum #1\expandafter \@firstoftwo
 \else \expandafter \@secondoftwo
 \fi
}%
\providecommand \@ifx [1]{%
 \ifx #1\expandafter \@firstoftwo
 \else \expandafter \@secondoftwo
 \fi
}%
\providecommand \natexlab [1]{#1}%
\providecommand \enquote  [1]{``#1''}%
\providecommand \bibnamefont  [1]{#1}%
\providecommand \bibfnamefont [1]{#1}%
\providecommand \citenamefont [1]{#1}%
\providecommand \href@noop [0]{\@secondoftwo}%
\providecommand \href [0]{\begingroup \@sanitize@url \@href}%
\providecommand \@href[1]{\@@startlink{#1}\@@href}%
\providecommand \@@href[1]{\endgroup#1\@@endlink}%
\providecommand \@sanitize@url [0]{\catcode `\\12\catcode `\$12\catcode
  `\&12\catcode `\#12\catcode `\^12\catcode `\_12\catcode `\%12\relax}%
\providecommand \@@startlink[1]{}%
\providecommand \@@endlink[0]{}%
\providecommand \url  [0]{\begingroup\@sanitize@url \@url }%
\providecommand \@url [1]{\endgroup\@href {#1}{\urlprefix }}%
\providecommand \urlprefix  [0]{URL }%
\providecommand \Eprint [0]{\href }%
\providecommand \doibase [0]{https://doi.org/}%
\providecommand \selectlanguage [0]{\@gobble}%
\providecommand \bibinfo  [0]{\@secondoftwo}%
\providecommand \bibfield  [0]{\@secondoftwo}%
\providecommand \translation [1]{[#1]}%
\providecommand \BibitemOpen [0]{}%
\providecommand \bibitemStop [0]{}%
\providecommand \bibitemNoStop [0]{.\EOS\space}%
\providecommand \EOS [0]{\spacefactor3000\relax}%
\providecommand \BibitemShut  [1]{\csname bibitem#1\endcsname}%
\let\auto@bib@innerbib\@empty
%</preamble>
\bibitem [{\citenamefont {Choi}\ \emph {et~al.}(2003)\citenamefont {Choi} \emph
  {et~al.}}]{Cho03}%
  \BibitemOpen
  \bibfield  {author} {\bibinfo {author} {\bibfnamefont {S.-K.}\ \bibnamefont
  {Choi}} \emph {et~al.} (\bibinfo {collaboration} {Belle Collaboration}),\
  }\bibfield  {title} {\bibinfo {title} {{Observation of a Narrow
  Charmoniumlike State in Exclusive \ensuremath{B^{\pm}\rightarrow
  K^{\pm}\pi^{+}\pi^{-}J/\psi} Decays}},\ }\href
  {https://doi.org/10.1103/PhysRevLett.91.262001} {\bibfield  {journal}
  {\bibinfo  {journal} {Phys. Rev. Lett.}\ }\textbf {\bibinfo {volume} {91}},\
  \bibinfo {pages} {262001} (\bibinfo {year} {2003})}\BibitemShut {NoStop}%
\bibitem [{\citenamefont {Zyla}\ \emph {et~al.}(2020)\citenamefont {Zyla} \emph
  {et~al.}}]{PDG20}%
  \BibitemOpen
  \bibfield  {author} {\bibinfo {author} {\bibfnamefont {P.~A.}\ \bibnamefont
  {Zyla}} \emph {et~al.} (\bibinfo {collaboration} {Particle Data Group}),\
  }\bibfield  {title} {\bibinfo {title} {{Review of Particle Physics}},\ }\href
  {https://doi.org/10.1093/ptep/ptaa104} {\bibfield  {journal} {\bibinfo
  {journal} {Prog. Theor. Exp. Phys.}\ }\textbf {\bibinfo {volume} {2020}},\
  \bibinfo {pages} {083C01} (\bibinfo {year} {2020})}\BibitemShut {NoStop}%
\bibitem [{\citenamefont {Eichten}\ \emph {et~al.}(1978)\citenamefont
  {Eichten}, \citenamefont {Gottfried}, \citenamefont {Kinoshita},
  \citenamefont {Lane},\ and\ \citenamefont {Yan}}]{Eic80}%
  \BibitemOpen
  \bibfield  {author} {\bibinfo {author} {\bibfnamefont {E.}~\bibnamefont
  {Eichten}}, \bibinfo {author} {\bibfnamefont {K.}~\bibnamefont {Gottfried}},
  \bibinfo {author} {\bibfnamefont {T.}~\bibnamefont {Kinoshita}}, \bibinfo
  {author} {\bibfnamefont {K.~D.}\ \bibnamefont {Lane}},\ and\ \bibinfo
  {author} {\bibfnamefont {T.~M.}\ \bibnamefont {Yan}},\ }\bibfield  {title}
  {\bibinfo {title} {{Charmonium: The model}},\ }\href
  {https://doi.org/10.1103/PhysRevD.17.3090} {\bibfield  {journal} {\bibinfo
  {journal} {Phys. Rev. D}\ }\textbf {\bibinfo {volume} {17}},\ \bibinfo
  {pages} {3090} (\bibinfo {year} {1978})}\BibitemShut {NoStop}%
\bibitem [{\citenamefont {Eichten}\ \emph {et~al.}(2004)\citenamefont
  {Eichten}, \citenamefont {Lane},\ and\ \citenamefont {Quigg}}]{Eic04}%
  \BibitemOpen
  \bibfield  {author} {\bibinfo {author} {\bibfnamefont {E.~J.}\ \bibnamefont
  {Eichten}}, \bibinfo {author} {\bibfnamefont {K.}~\bibnamefont {Lane}},\ and\
  \bibinfo {author} {\bibfnamefont {C.}~\bibnamefont {Quigg}},\ }\bibfield
  {title} {\bibinfo {title} {{Charmonium levels near threshold and the narrow
  state \ensuremath{X(3872)\rightarrow\pi^{+}\pi^{-}J/\psi}}},\ }\href
  {https://doi.org/10.1103/PhysRevD.69.094019} {\bibfield  {journal} {\bibinfo
  {journal} {Phys. Rev. D}\ }\textbf {\bibinfo {volume} {69}},\ \bibinfo
  {pages} {094019} (\bibinfo {year} {2004})}\BibitemShut {NoStop}%
\bibitem [{\citenamefont {Godfrey}\ and\ \citenamefont {Isgur}(1985)}]{GI85}%
  \BibitemOpen
  \bibfield  {author} {\bibinfo {author} {\bibfnamefont {S.}~\bibnamefont
  {Godfrey}}\ and\ \bibinfo {author} {\bibfnamefont {N.}~\bibnamefont
  {Isgur}},\ }\bibfield  {title} {\bibinfo {title} {{Mesons in a relativized
  quark model with chromodynamics}},\ }\href
  {https://doi.org/10.1103/PhysRevD.32.189} {\bibfield  {journal} {\bibinfo
  {journal} {Phys. Rev. D}\ }\textbf {\bibinfo {volume} {32}},\ \bibinfo
  {pages} {189} (\bibinfo {year} {1985})}\BibitemShut {NoStop}%
\bibitem [{\citenamefont {Barnes}\ and\ \citenamefont {Swanson}(2008)}]{Bar08}%
  \BibitemOpen
  \bibfield  {author} {\bibinfo {author} {\bibfnamefont {T.}~\bibnamefont
  {Barnes}}\ and\ \bibinfo {author} {\bibfnamefont {E.~S.}\ \bibnamefont
  {Swanson}},\ }\bibfield  {title} {\bibinfo {title} {{Hadron loops: General
  theorems and application to charmonium}},\ }\href
  {https://doi.org/10.1103/PhysRevC.77.055206} {\bibfield  {journal} {\bibinfo
  {journal} {Phys. Rev. C}\ }\textbf {\bibinfo {volume} {77}},\ \bibinfo
  {pages} {055206} (\bibinfo {year} {2008})}\BibitemShut {NoStop}%
\bibitem [{\citenamefont {Ferretti}\ and\ \citenamefont
  {Santopinto}(2019)}]{Fer19}%
  \BibitemOpen
  \bibfield  {author} {\bibinfo {author} {\bibfnamefont {J.}~\bibnamefont
  {Ferretti}}\ and\ \bibinfo {author} {\bibfnamefont {E.}~\bibnamefont
  {Santopinto}},\ }\bibfield  {title} {\bibinfo {title} {{Threshold corrections
  of \ensuremath{\chi_c(2P)} and \ensuremath{\chi_b(3P)} states and
  \ensuremath{J/\psi\rho} and \ensuremath{J/\psi\omega} transitions of the
  \ensuremath{X(3872)} in a coupled-channel model}},\ }\href
  {https://doi.org/10.1016/j.physletb.2018.12.052} {\bibfield  {journal}
  {\bibinfo  {journal} {Phys. Lett. B}\ }\textbf {\bibinfo {volume} {789}},\
  \bibinfo {pages} {550} (\bibinfo {year} {2019})}\BibitemShut {NoStop}%
\bibitem [{\citenamefont {Chen}\ \emph {et~al.}(2016)\citenamefont {Chen},
  \citenamefont {Chen}, \citenamefont {Liu},\ and\ \citenamefont {Zhu}}]{Ch16}%
  \BibitemOpen
  \bibfield  {author} {\bibinfo {author} {\bibfnamefont {H.-X.}\ \bibnamefont
  {Chen}}, \bibinfo {author} {\bibfnamefont {W.}~\bibnamefont {Chen}}, \bibinfo
  {author} {\bibfnamefont {X.}~\bibnamefont {Liu}},\ and\ \bibinfo {author}
  {\bibfnamefont {S.-L.}\ \bibnamefont {Zhu}},\ }\bibfield  {title} {\bibinfo
  {title} {{The hidden-charm pentaquark and tetraquark states}},\ }\href
  {https://doi.org/10.1016/j.physrep.2016.05.004} {\bibfield  {journal}
  {\bibinfo  {journal} {Phys. Rept.}\ }\textbf {\bibinfo {volume} {639}},\
  \bibinfo {pages} {1} (\bibinfo {year} {2016})}\BibitemShut {NoStop}%
\bibitem [{\citenamefont {Lebed}\ \emph {et~al.}(2017)\citenamefont {Lebed},
  \citenamefont {Mitchell},\ and\ \citenamefont {Swanson}}]{Leb17}%
  \BibitemOpen
  \bibfield  {author} {\bibinfo {author} {\bibfnamefont {R.~F.}\ \bibnamefont
  {Lebed}}, \bibinfo {author} {\bibfnamefont {R.~E.}\ \bibnamefont
  {Mitchell}},\ and\ \bibinfo {author} {\bibfnamefont {E.~S.}\ \bibnamefont
  {Swanson}},\ }\bibfield  {title} {\bibinfo {title} {{Heavy-quark QCD
  exotica}},\ }\href {https://doi.org/10.1016/j.ppnp.2016.11.003} {\bibfield
  {journal} {\bibinfo  {journal} {Prog. Part. Nucl. Phys.}\ }\textbf {\bibinfo
  {volume} {93}},\ \bibinfo {pages} {143} (\bibinfo {year} {2017})}\BibitemShut
  {NoStop}%
\bibitem [{\citenamefont {Guo}\ \emph {et~al.}(2018)\citenamefont {Guo},
  \citenamefont {Hanhart}, \citenamefont {Mei\ss{}ner}, \citenamefont {Wang},
  \citenamefont {Zhao},\ and\ \citenamefont {Zou}}]{Guo17}%
  \BibitemOpen
  \bibfield  {author} {\bibinfo {author} {\bibfnamefont {F.-K.}\ \bibnamefont
  {Guo}}, \bibinfo {author} {\bibfnamefont {C.}~\bibnamefont {Hanhart}},
  \bibinfo {author} {\bibfnamefont {U.-G.}\ \bibnamefont {Mei\ss{}ner}},
  \bibinfo {author} {\bibfnamefont {Q.}~\bibnamefont {Wang}}, \bibinfo {author}
  {\bibfnamefont {Q.}~\bibnamefont {Zhao}},\ and\ \bibinfo {author}
  {\bibfnamefont {B.-S.}\ \bibnamefont {Zou}},\ }\bibfield  {title} {\bibinfo
  {title} {{Hadronic molecules}},\ }\href
  {https://doi.org/10.1103/RevModPhys.90.015004} {\bibfield  {journal}
  {\bibinfo  {journal} {Rev. Mod. Phys.}\ }\textbf {\bibinfo {volume} {90}},\
  \bibinfo {pages} {015004} (\bibinfo {year} {2018})}\BibitemShut {NoStop}%
\bibitem [{\citenamefont {Esposito}\ \emph {et~al.}(2017)\citenamefont
  {Esposito}, \citenamefont {Pilloni},\ and\ \citenamefont {Polosa}}]{Esp17}%
  \BibitemOpen
  \bibfield  {author} {\bibinfo {author} {\bibfnamefont {A.}~\bibnamefont
  {Esposito}}, \bibinfo {author} {\bibfnamefont {A.}~\bibnamefont {Pilloni}},\
  and\ \bibinfo {author} {\bibfnamefont {A.}~\bibnamefont {Polosa}},\
  }\bibfield  {title} {\bibinfo {title} {{Multiquark resonances}},\ }\href
  {https://doi.org/10.1016/j.physrep.2016.11.002} {\bibfield  {journal}
  {\bibinfo  {journal} {Phys. Rept.}\ }\textbf {\bibinfo {volume} {668}},\
  \bibinfo {pages} {1} (\bibinfo {year} {2017})}\BibitemShut {NoStop}%
\bibitem [{\citenamefont {Juge}\ \emph {et~al.}(1999)\citenamefont {Juge},
  \citenamefont {Kuti},\ and\ \citenamefont {Morningstar}}]{Jug99}%
  \BibitemOpen
  \bibfield  {author} {\bibinfo {author} {\bibfnamefont {K.~J.}\ \bibnamefont
  {Juge}}, \bibinfo {author} {\bibfnamefont {J.}~\bibnamefont {Kuti}},\ and\
  \bibinfo {author} {\bibfnamefont {C.~J.}\ \bibnamefont {Morningstar}},\
  }\bibfield  {title} {\bibinfo {title} {{Ab Initio Study of Hybrid
  \ensuremath{\overline{b}gb} Mesons}},\ }\href
  {https://doi.org/10.1103/PhysRevLett.82.4400} {\bibfield  {journal} {\bibinfo
   {journal} {Phys. Rev. Lett.}\ }\textbf {\bibinfo {volume} {82}},\ \bibinfo
  {pages} {4400} (\bibinfo {year} {1999})}\BibitemShut {NoStop}%
\bibitem [{\citenamefont {Brambilla}\ \emph {et~al.}(2019)\citenamefont
  {Brambilla}, \citenamefont {Eidelman}, \citenamefont {Hanhart}, \citenamefont
  {Nefediev}, \citenamefont {Shen}, \citenamefont {Thomas}, \citenamefont
  {Vairo},\ and\ \citenamefont {Yuan}}]{Bra20}%
  \BibitemOpen
  \bibfield  {author} {\bibinfo {author} {\bibfnamefont {N.}~\bibnamefont
  {Brambilla}}, \bibinfo {author} {\bibfnamefont {S.}~\bibnamefont {Eidelman}},
  \bibinfo {author} {\bibfnamefont {C.}~\bibnamefont {Hanhart}}, \bibinfo
  {author} {\bibfnamefont {A.}~\bibnamefont {Nefediev}}, \bibinfo {author}
  {\bibfnamefont {C.-P.}\ \bibnamefont {Shen}}, \bibinfo {author}
  {\bibfnamefont {C.~E.}\ \bibnamefont {Thomas}}, \bibinfo {author}
  {\bibfnamefont {A.}~\bibnamefont {Vairo}},\ and\ \bibinfo {author}
  {\bibfnamefont {C.-Z.}\ \bibnamefont {Yuan}},\ }\bibfield  {title} {\bibinfo
  {title} {{The \ensuremath{XYZ} states: experimental and theoretical status
  and perspectives}},\ }\Eprint {https://arxiv.org/abs/1907.07583}
  {arXiv:1907.07583 [hep-ex]}  (\bibinfo {year} {2019})\BibitemShut {NoStop}%
\bibitem [{\citenamefont {Bali}(2001)}]{Bal01}%
  \BibitemOpen
  \bibfield  {author} {\bibinfo {author} {\bibfnamefont {G.~S.}\ \bibnamefont
  {Bali}},\ }\bibfield  {title} {\bibinfo {title} {{QCD forces and heavy quark
  bound states}},\ }\href {https://doi.org/10.1016/S0370-1573(00)00079-X}
  {\bibfield  {journal} {\bibinfo  {journal} {Phys. Rept.}\ }\textbf {\bibinfo
  {volume} {343}},\ \bibinfo {pages} {1} (\bibinfo {year} {2001})}\BibitemShut
  {NoStop}%
\bibitem [{\citenamefont {Braaten}\ \emph {et~al.}(2014)\citenamefont
  {Braaten}, \citenamefont {Langmack},\ and\ \citenamefont {Smith}}]{Braa14}%
  \BibitemOpen
  \bibfield  {author} {\bibinfo {author} {\bibfnamefont {E.}~\bibnamefont
  {Braaten}}, \bibinfo {author} {\bibfnamefont {C.}~\bibnamefont {Langmack}},\
  and\ \bibinfo {author} {\bibfnamefont {D.~H.}\ \bibnamefont {Smith}},\
  }\bibfield  {title} {\bibinfo {title} {{Born-Oppenheimer approximation for
  the \ensuremath{XYZ} mesons}},\ }\href
  {https://doi.org/10.1103/PhysRevD.90.014044} {\bibfield  {journal} {\bibinfo
  {journal} {Phys. Rev. D}\ }\textbf {\bibinfo {volume} {90}},\ \bibinfo
  {pages} {014044} (\bibinfo {year} {2014})}\BibitemShut {NoStop}%
\bibitem [{\citenamefont {Gonz{\'{a}}lez}(2014)}]{Gon14}%
  \BibitemOpen
  \bibfield  {author} {\bibinfo {author} {\bibfnamefont {P.}~\bibnamefont
  {Gonz{\'{a}}lez}},\ }\bibfield  {title} {\bibinfo {title} {{Generalized
  screened potential model}},\ }\href
  {https://doi.org/10.1088/0954-3899/41/9/095001} {\bibfield  {journal}
  {\bibinfo  {journal} {J. Phys. G}\ }\textbf {\bibinfo {volume} {41}},\
  \bibinfo {pages} {095001} (\bibinfo {year} {2014})}\BibitemShut {NoStop}%
\bibitem [{\citenamefont {Gonz\'alez}(2015)}]{Gon15}%
  \BibitemOpen
  \bibfield  {author} {\bibinfo {author} {\bibfnamefont {P.}~\bibnamefont
  {Gonz\'alez}},\ }\bibfield  {title} {\bibinfo {title} {{Charmonium
  description from a generalized screened potential model}},\ }\href
  {https://doi.org/10.1103/PhysRevD.92.014017} {\bibfield  {journal} {\bibinfo
  {journal} {Phys. Rev. D}\ }\textbf {\bibinfo {volume} {92}},\ \bibinfo
  {pages} {014017} (\bibinfo {year} {2015})}\BibitemShut {NoStop}%
\bibitem [{\citenamefont {Bruschini}\ and\ \citenamefont
  {Gonz\'alez}(2019)}]{Gon19}%
  \BibitemOpen
  \bibfield  {author} {\bibinfo {author} {\bibfnamefont {R.}~\bibnamefont
  {Bruschini}}\ and\ \bibinfo {author} {\bibfnamefont {P.}~\bibnamefont
  {Gonz\'alez}},\ }\bibfield  {title} {\bibinfo {title} {{Quark model
  description of \ensuremath{\psi(4260)}}},\ }\href
  {https://doi.org/10.1103/PhysRevC.99.045205} {\bibfield  {journal} {\bibinfo
  {journal} {Phys. Rev. C}\ }\textbf {\bibinfo {volume} {99}},\ \bibinfo
  {pages} {045205} (\bibinfo {year} {2019})}\BibitemShut {NoStop}%
\bibitem [{\citenamefont {Bali}\ \emph {et~al.}(2005)\citenamefont {Bali},
  \citenamefont {Neff}, \citenamefont {D\"ussel}, \citenamefont {Lippert},\
  and\ \citenamefont {Schilling}}]{Bal05}%
  \BibitemOpen
  \bibfield  {author} {\bibinfo {author} {\bibfnamefont {G.~S.}\ \bibnamefont
  {Bali}}, \bibinfo {author} {\bibfnamefont {H.}~\bibnamefont {Neff}}, \bibinfo
  {author} {\bibfnamefont {T.}~\bibnamefont {D\"ussel}}, \bibinfo {author}
  {\bibfnamefont {T.}~\bibnamefont {Lippert}},\ and\ \bibinfo {author}
  {\bibfnamefont {K.}~\bibnamefont {Schilling}} (\bibinfo {collaboration}
  {SESAM Collaboration}),\ }\bibfield  {title} {\bibinfo {title} {{Observation
  of string breaking in QCD}},\ }\href
  {https://doi.org/10.1103/PhysRevD.71.114513} {\bibfield  {journal} {\bibinfo
  {journal} {Phys. Rev. D}\ }\textbf {\bibinfo {volume} {71}},\ \bibinfo
  {pages} {114513} (\bibinfo {year} {2005})}\BibitemShut {NoStop}%
\bibitem [{\citenamefont {Bulava}\ \emph {et~al.}(2019)\citenamefont {Bulava},
  \citenamefont {Hörz}, \citenamefont {Knechtli}, \citenamefont {Koch},
  \citenamefont {Moir}, \citenamefont {Morningstar},\ and\ \citenamefont
  {Peardon}}]{Bul19}%
  \BibitemOpen
  \bibfield  {author} {\bibinfo {author} {\bibfnamefont {J.}~\bibnamefont
  {Bulava}}, \bibinfo {author} {\bibfnamefont {B.}~\bibnamefont {Hörz}},
  \bibinfo {author} {\bibfnamefont {F.}~\bibnamefont {Knechtli}}, \bibinfo
  {author} {\bibfnamefont {V.}~\bibnamefont {Koch}}, \bibinfo {author}
  {\bibfnamefont {G.}~\bibnamefont {Moir}}, \bibinfo {author} {\bibfnamefont
  {C.}~\bibnamefont {Morningstar}},\ and\ \bibinfo {author} {\bibfnamefont
  {M.}~\bibnamefont {Peardon}},\ }\bibfield  {title} {\bibinfo {title} {{String
  breaking by light and strange quarks in QCD}},\ }\href
  {https://doi.org/10.1016/j.physletb.2019.05.018} {\bibfield  {journal}
  {\bibinfo  {journal} {Phys. Lett. B}\ }\textbf {\bibinfo {volume} {793}},\
  \bibinfo {pages} {493} (\bibinfo {year} {2019})}\BibitemShut {NoStop}%
\bibitem [{\citenamefont {Baer}(2006)}]{Bae06}%
  \BibitemOpen
  \bibfield  {author} {\bibinfo {author} {\bibfnamefont {M.}~\bibnamefont
  {Baer}},\ }\href@noop {} {\emph {\bibinfo {title} {{Beyond Born-Oppenheimer:
  electronic nonadiabatic coupling terms and conical intersections}}}}\
  (\bibinfo  {publisher} {John Wiley \& Sons},\ \bibinfo {year}
  {2006})\BibitemShut {NoStop}%
\bibitem [{\citenamefont {Born}\ and\ \citenamefont
  {Oppenheimer}(1927)}]{Bo27}%
  \BibitemOpen
  \bibfield  {author} {\bibinfo {author} {\bibfnamefont {M.}~\bibnamefont
  {Born}}\ and\ \bibinfo {author} {\bibfnamefont {R.}~\bibnamefont
  {Oppenheimer}},\ }\bibfield  {title} {\bibinfo {title} {{Zur Quantentheorie
  der Molekeln}},\ }\href {https://doi.org/10.1002/andp.19273892002} {\bibfield
   {journal} {\bibinfo  {journal} {Ann. Phys. (Berl.)}\ }\textbf {\bibinfo
  {volume} {389}},\ \bibinfo {pages} {457} (\bibinfo {year}
  {1927})}\BibitemShut {NoStop}%
\bibitem [{\citenamefont {Bicudo}\ \emph {et~al.}(2020)\citenamefont {Bicudo},
  \citenamefont {Cardoso}, \citenamefont {Cardoso},\ and\ \citenamefont
  {Wagner}}]{Bic20}%
  \BibitemOpen
  \bibfield  {author} {\bibinfo {author} {\bibfnamefont {P.}~\bibnamefont
  {Bicudo}}, \bibinfo {author} {\bibfnamefont {M.}~\bibnamefont {Cardoso}},
  \bibinfo {author} {\bibfnamefont {N.}~\bibnamefont {Cardoso}},\ and\ \bibinfo
  {author} {\bibfnamefont {M.}~\bibnamefont {Wagner}},\ }\bibfield  {title}
  {\bibinfo {title} {{Bottomonium resonances with \ensuremath{I=0} from lattice
  QCD correlation functions with static and light quarks}},\ }\href
  {https://doi.org/10.1103/PhysRevD.101.034503} {\bibfield  {journal} {\bibinfo
   {journal} {Phys. Rev. D}\ }\textbf {\bibinfo {volume} {101}},\ \bibinfo
  {pages} {034503} (\bibinfo {year} {2020})}\BibitemShut {NoStop}%
\bibitem [{\citenamefont {Eichten}\ and\ \citenamefont {Quigg}(1994)}]{Eic94}%
  \BibitemOpen
  \bibfield  {author} {\bibinfo {author} {\bibfnamefont {E.~J.}\ \bibnamefont
  {Eichten}}\ and\ \bibinfo {author} {\bibfnamefont {C.}~\bibnamefont
  {Quigg}},\ }\bibfield  {title} {\bibinfo {title} {{Mesons with beauty and
  charm: Spectroscopy}},\ }\href {https://doi.org/10.1103/PhysRevD.49.5845}
  {\bibfield  {journal} {\bibinfo  {journal} {Phys. Rev. D}\ }\textbf {\bibinfo
  {volume} {49}},\ \bibinfo {pages} {5845} (\bibinfo {year}
  {1994})}\BibitemShut {NoStop}%
\bibitem [{\citenamefont {Bruschini}\ and\ \citenamefont
  {Gonz\'alez}(2020)}]{Bru20}%
  \BibitemOpen
  \bibfield  {author} {\bibinfo {author} {\bibfnamefont {R.}~\bibnamefont
  {Bruschini}}\ and\ \bibinfo {author} {\bibfnamefont {P.}~\bibnamefont
  {Gonz\'alez}},\ }\bibfield  {title} {\bibinfo {title} {{Radiative decays in
  charmonium beyond the \ensuremath{p/m} approximation}},\ }\href
  {https://doi.org/10.1103/PhysRevD.101.014027} {\bibfield  {journal} {\bibinfo
   {journal} {Phys. Rev. D}\ }\textbf {\bibinfo {volume} {101}},\ \bibinfo
  {pages} {014027} (\bibinfo {year} {2020})}\BibitemShut {NoStop}%
\bibitem [{\citenamefont {Badalian}\ \emph {et~al.}(2009)\citenamefont
  {Badalian}, \citenamefont {Bakker},\ and\ \citenamefont {Danilkin}}]{Bad09}%
  \BibitemOpen
  \bibfield  {author} {\bibinfo {author} {\bibfnamefont {A.}~\bibnamefont
  {Badalian}}, \bibinfo {author} {\bibfnamefont {B.}~\bibnamefont {Bakker}},\
  and\ \bibinfo {author} {\bibfnamefont {I.}~\bibnamefont {Danilkin}},\
  }\bibfield  {title} {\bibinfo {title} {{The \ensuremath{S}-\ensuremath{D}
  mixing and di-electron widths of higher charmonium \ensuremath{1^{--}}
  states}},\ }\href {https://doi.org/10.1134/S1063778809040085} {\bibfield
  {journal} {\bibinfo  {journal} {Phys. Atom. Nucl.}\ }\textbf {\bibinfo
  {volume} {72}},\ \bibinfo {pages} {638} (\bibinfo {year} {2009})}\BibitemShut
  {NoStop}%
\end{thebibliography}%

\end{document}